\documentclass[twocolumn,superscriptaddress,prl]{revtex4-2}
\usepackage[utf8]{inputenc}
\usepackage{amsfonts}
\usepackage{graphicx}
\usepackage{color}
\usepackage{soul} 
\usepackage{amsmath}
\usepackage{hyperref}
\usepackage{braket}
\usepackage{mathtools}
\usepackage{lipsum}
\hypersetup{colorlinks=true, citecolor=cyan, urlcolor=blue, linkcolor=blue}
\usepackage{mathrsfs}

\newcommand{\RMP}[4]{\textit{#1}, Rev. Mod. Phys. \textbf{#2}, #3 (#4)}

\newcommand{\PRL}[4]{\textit{#1}, Phys. Rev. Lett. \textbf{#2}, #3 (#4)}
\newcommand{\PRR}[4]{\textit{#1}, Phys. Rev. Research. \textbf{#2}, #3 (#4)}
\newcommand{\PRA}[4]{\textit{#1}, Phys. Rev. A \textbf{#2}, #3 (#4)}
\newcommand{\PRB}[4]{\textit{#1}, Phys. Rev. B \textbf{#2}, #3 (#4)}
\newcommand{\PRE}[4]{\textit{#1}, Phys. Rev. E \textbf{#2}, #3 (#4)}
\newcommand{\PRApplied}[4]{\textit{#1}, Phys. Rev. Appl. \textbf{#2}, #3 (#4)}
\newcommand{\PRX}[4]{\textit{#1}, Phys. Rev. X \textbf{#2}, #3 (#4)}
\newcommand{\PRXQ}[4]{\textit{#1}, PRX Quantum \textbf{#2}, #3 (#4)}

\newcommand{\NComm}[4]{\textit{#1}, Nat. Comm \textbf{#2}, #3 (#4)}
\newcommand{\CommPhys}[4]{\textit{#1}, Comm. Phys \textbf{#2}, #3 (#4)}
\newcommand{\SciAdv}[4]{\textit{#1}, Sci. Adv \textbf{#2}, #3 (#4)}

\newcommand{\JLT}[4]{\textit{#1}, J. Low. Temp. Phys. \textbf{#2}, #3 (#4)}

\begin{document}
\title{
Evidence for unexpectedly low quasiparticle generation rates \\ across Josephson junctions of driven superconducting qubits
}
\author{Byoung-moo Ann}
\email{byoungmoo.ann@gmail.com}
\affiliation{Quantum Technology Institute, Korea Research Institute of Standards and Science, 34113 Daejeon, South Korea}
\author{Sang-Jun Choi}
\affiliation{Department of Physics Education, Kongju National University, Gongju 32588, South Korea} 
\author{Hee Chul Park}
\affiliation{Department of Physics, Pukyong National University, Busan 48513, South Korea} 
\author{Sercan Deve}
\affiliation{Kavli Institute of Nanoscience, Delft University of Technology, 2628 CJ Delft, The Netherlands} 
\author{Robin Dekker}
\affiliation{Kavli Institute of Nanoscience, Delft University of Technology, 2628 CJ Delft, The Netherlands} 
\author{Gary A. Steele}
\affiliation{Kavli Institute of Nanoscience, Delft University of Technology, 2628 CJ Delft, The Netherlands} 
\author{Jaseung Ku}
\affiliation{Quantum Technology Institute, Korea Research Institute of Standards and Science, 34113 Daejeon, South Korea}
\author{Seung-Bo Shim}
\affiliation{Quantum Technology Institute, Korea Research Institute of Standards and Science, 34113 Daejeon, South Korea}
\author{Junho Suh}
\email{junhosuh@postech.ac.kr}
\affiliation{Quantum Technology Institute, Korea Research Institute of Standards and Science, 34113 Daejeon, South Korea}
\affiliation{Department of Physics, Pohang University of Science and Technology (POSTECH), Pohang,  South Korea}

\date{\today}

\begin{abstract}

Recent studies find that even drives far below the superconducting gap frequency may cause drive-induced quasiparticle generation (QPG) across Josephson junctions (JJs) of superconducting qubits (SCQs), posing a serious concern for fault-tolerant superconducting quantum computing (FTSQC).
Nonetheless, quantitative experimental estimation on QPG rates has remained vague.
Here, we investigate QPG using strongly driven SCQs, reaching qubit drive amplitudes up to $2\pi\times$300 GHz by applying intense drive fields through the readout resonators.
The resonator nonlinear responses enable quantification of the energy loss at SCQs, including the contribution from QPG.
Surprisingly, the estimated total energy loss rates are far lower than those expected by the Floquet-Markov formalism with QPG as the sole loss mechanism.
Meanwhile, calculations that incorporate high-frequency cutoffs (HFCs) in the QPG conductance at approximately 17–20 GHz effectively explain the experimental observations.
These results suggest limitations in either the QPG conductance model or the Markovian treatment of the QPG processes. 
Both possibilities possess crucial implications for handling QPG problems toward FTSQC and for a more deeper understanding of Josephson junctions.

%

\end{abstract}

\maketitle


Circuit quantum electrodynamics (cQED) with superconducting qubits (SCQs) and readout resonators is one of the most promising platform for implementing quantum technologies \cite{cQED}.
Applying time-periodic drives on SCQs is general technique, which has been investigated in various contexts \cite{driven-system-1,driven-system-2,driven-system-3,driven-system-4,driven-system-5,driven-system-6,driven-system-7,driven-system-8,driven-system-9,driven-system-10,driven-system-11,driven-system-12,driven-system-13,driven-system-14}.
In the past few years, the drive-induced unwanted state transitions (DUST) to higher energy levels have been intensively studies as a limiting factor on the strength of drives, which ultimately limit performances of SCQs \cite{DUST-1, DUST-2, DUST-3, DUST-4}. 
%
%
Several mitigation schemes has been suggested and demonstrated \cite{HFR, Chapple-PRApplied-2025, Mori-2025, Li-2025}.

More recently, theoretical studies have suggested that microwave drives with frequencies far below the superconducting gap can still induce quasiparticle generation (QPG) across the Josephson junctions (JJs) \cite{QPG-1, QPG-2}, which poses further challenges.
Moreover, drive-induced QPG becomes more severe at higher drive frequencies, contradicting the high-frequency readout strategy \cite{HFR}, which is largely attended to mitigate DUST during qubit readout.
Thus, the future progress of superconducting quantum computing will hinge on overcoming the constraints simultaneously imposed by DUST and drive-induced QPG.
Despite such significance, quantitative experimental study on drive-induced QPG has been largely unexplored.

In this article, we experimentally investigate drive-induced QPG using superconducting transmon qubits \cite{Koch-PRA-2007}.
When a readout resonator coupled to a transmon is strongly probed, a prominent feature in its response is the appearance of the bare resonator frequency.
We observe that the dissipation associated with this bare response decreases markedly as the probe power increases, over probe photon numbers ranging from 400 to $5\times10^5$. 
This nonlinear dissipation (ND) cannot be explained by existing ND models.
This ND allows us to extract the energy loss occurring at the transmon, and any loss arising from QPG should be contained within this.

Comparing the experimentally observed energy loss with numerical results obtained using the Floquet–Markov formalism leads to a striking conclusion.
Even under the assumption that all dissipation originates from QPG, the measured loss is far smaller than predicted by numerical simulations using the ideal QPG model. 
We obtain the same evidence from multiple transmons.
Meanwhile, including high-frequency cutoffs (HFCs) around 17–20 GHz to the QPG conductance and preserving Markovic approximation effectively solves this discrepancy.
We provide fundamental and practical implications of our results in the last section.
Therein, we also suggest other possible methods of investigating drive-induced QPG and present expected technical difficulties.

\begin{figure*}
    \centering
    \includegraphics[width=0.8\linewidth]{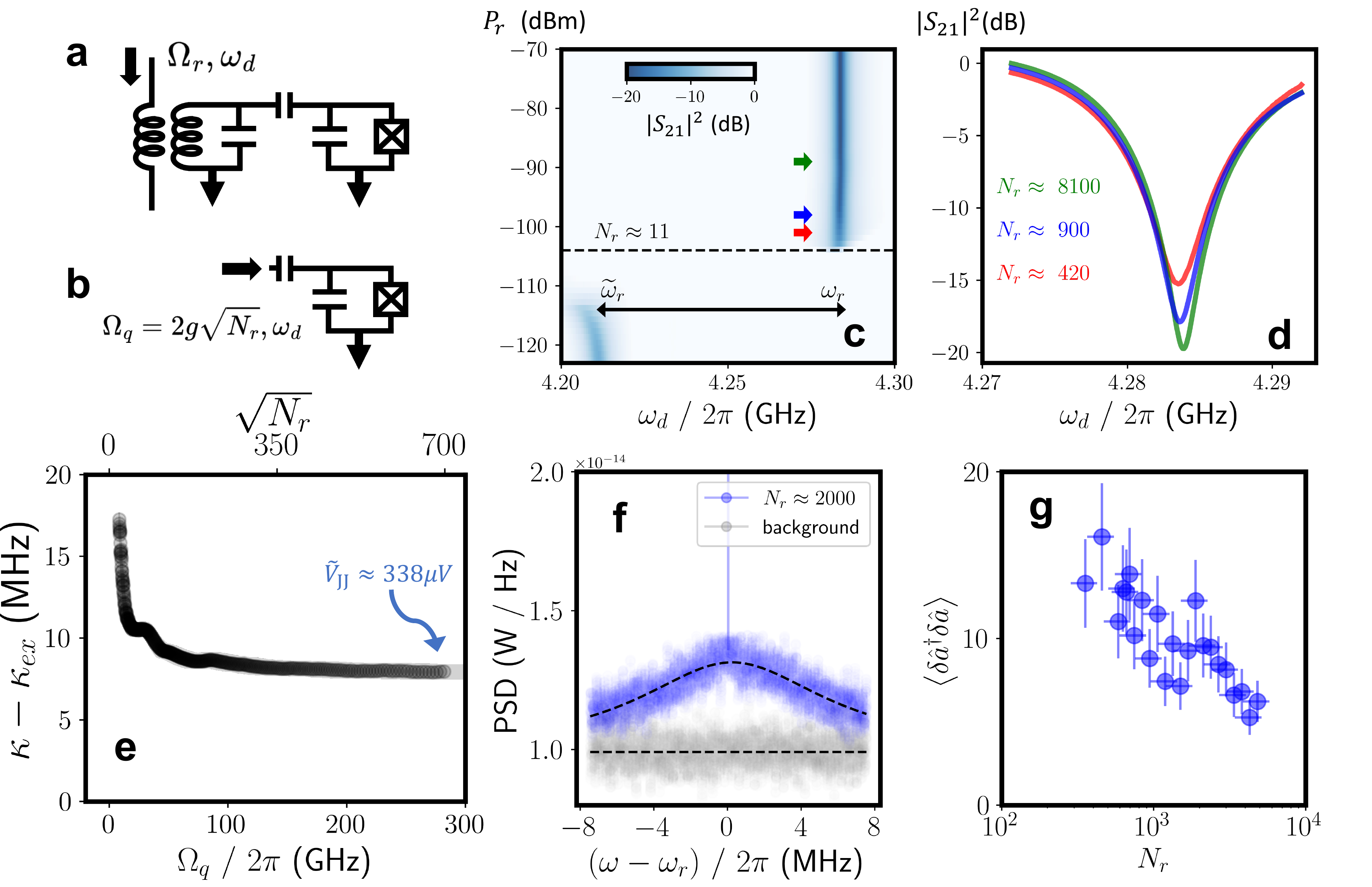}
    \caption{\textbf{Nonlinear dissipation in the bare resonator response.} \textbf{(a)} A transmon is coupled to a resonator mode of the bare frequency $\omega_r$. The arrow indicates a resonator probe of frequency $\omega_d$ and amplitude $\Omega_r$. 
    \textbf{(b)} Equivalent schematic in the semi-classical approximation. The arrow indicates an effective drive on the transmon. Here, $g$ and $N_r$ are transmon–resonator coupling strength and resonator mean photon number, respectively. 
    \textbf{(c)} Observed resonator transmission ($S_{21}$) with respect to resonator probe power $P_r$. The collapse of dressed resonator frequencies ($\widetilde{\omega}_r$) to the bare one ($\omega_r$) is identified (white arrow). The dashed line indicates where the bare resonator response appears. \textbf{(d)} Cross-sections corresponding to red, blue, and green arrows in (c) near $\omega_r$. Corresponding resonator photon numbers for resonant drives are presented beside. \textbf{(e)} Extracted resonator dissipation rates $\kappa$ (dots) for resonant drives neglecting resonator inhomogeneous broadening. $\widetilde{V}_{\textup{JJ}}$ refers to an estimated mean amplitude of voltage across the JJ in the transmon.
    The gray area represents scaling errors in the $x$-axes considering $\pm$20\% calibration error in $P_r$.
    \textbf{(f)} Power spectrum density of the resonator output when $N_r\approx2000$ (blue), background (gray), and fits (dashed lines). The sharp spike at the center is the transmitted probe. The broad peak corresponds to the spectrum of $\delta\hat{a}$. \textbf{(g)} Estimated $\left<\delta\hat{a}^\dagger\delta\hat{a} \right>$ versus $N_r$. Error bars represent scaling errors assuming $\pm$20\% calibration errors in $\left<\delta\hat{a}^\dagger\delta\hat{a} \right>$ and $N_r$. Statistical errors are negligible in all the cases. }
    \label{fig:1}
\end{figure*}

\section*{Nonlinear dissipation in the bare resonator response}
This section explains ND of readout resonators coupled to transmons in the bare frequency responses.
Fig.~\ref{fig:1}(a) provides a brief overview of our experiment.
The circuit consists of a dispersively coupled transmon and readout resonator. The transmon is based on a Al/AlOx/Al Josephson junction (JJ) fabricated based on the standard procedure. The other parts of the device are comprised of NbTiN film \cite{SRON}. The resonator is probed through a inductively side-coupled feedline with amplitude $\Omega_r$ and frequency $\omega_d$. We measure the transmission of the resonator probe. The measurement is performed at approximately 10 mK. $\textbf{Methods}$ provide explanations on the devices and experimental setup.

The Hamiltonian of the system reads
\begin{equation}
\begin{split}
\label{eq1}
    \hat{H}(t) & =  4E_C(\hat{n} - n_g)^2 - E_J\cos\hat{\varphi} + \hbar\omega_r\hat{a}^\dagger\hat{a} - i\hbar g\hat{n}(\hat{a} - \hat{a}^\dagger) \\ & + \Omega_r (\hat{a}+\hat{a}^{\dagger}) \sin \omega_d t.
\end{split}
\end{equation}
Here, $\hat{a}$, $\hat{n}$, and $\hat{\varphi}$ are resonator field, transmon Cooper-pair number, superconducting phase operators, respectively. $E_C=h\cdot259$ MHz, $E_J=h\cdot14.24$ GHz, $\omega_r/2\pi=4.284$ GHz, $g/2\pi=231$ MHz, and $n_g$ are charging, Josephson energies, bare resonator frequency, transmon–resonator coupling rate, and Cooper-pair number offset, respectively.
We can expand the resonator field as $\hat{a}=\left< \hat{a} \right> + {\delta \hat{a}}$, the mean of the resonator field and the fluctuations. Neglecting the latter amounts to the semi-classical approximation, which reduces $\hat{H}(t)$ to 
\begin{equation}
\begin{split}
\label{eq2}
    \hat{H}_q(t) & =  4E_C(\hat{n} - n_g)^2 - E_J\cos\hat{\varphi} + \hbar\Omega_q\hat{n} \cos \omega_d t,
\end{split}
\end{equation}
dropping the resonator-related operators. Fig.~\ref{fig:1}(b) describes an effectively equivalent system. Here, $\Omega_q=2g\sqrt{N_r}$ and $N_r=|\left\langle \hat{a} \right\rangle|^2$. See \textbf{Supplementary Note 1 and 7} for details on the derivation and device parameters.

Fig.~\ref{fig:1}(c) shows the measured resonator transmission ($S_{21}$) with respect to resonator probe power $P_r$. The shift from the dressed to bare resonator frequency ($\widetilde{\omega}_r \rightarrow \omega_r$) is identified. 
Across various studies \cite{driven-system-16, driven-system-17, driven-system-18, driven-system-19,driven-system-20,driven-system-21,driven-system-22}, this phenomenon has ultimately been attributed that the chaotic Floquet modes, dominated by unconfined states, largely participate in the transmon dynamics.

Although the bare resonator response appears around $N_r\approx11$ (dashed line), the data with $N_r \lesssim 400$ shows temporally unstable responses around $\omega_d$. Consequently, extracting accurate information is unfeasible.
Temporally stable responses are observed for $N_r \gtrsim 400$, and thus, we take only the data with $N_r \gtrsim 400$ into account hereinafter.
Fig.~\ref{fig:1}(d) presents some $|S_{21}|^2$ and corresponding $N_r$.
Neglecting resonator inhomogeneous broadening, $\kappa = \kappa_{ex} / (1-\min(|S_{21}|))$ and $N_r=\frac{2P_r}{\hbar\omega_d}\frac{\kappa_{ex}}{\kappa^2}$ hold for $\omega_d$ corresponding to $\min(|S_{21}|)$. $\kappa$ ($\kappa_{ex}$) is the total (external) resonator dissipation, and $\kappa_{ex}/2\pi=15.586$ MHz.
Noticeable changes in $S_{21}$ for different $P_r$ indicate nonlinear dissipation.

Fig.~\ref{fig:1}(e) present $\kappa-\kappa_{ex}$ extracted from $|S_{21}|^2$ (dark circles).
$\kappa$ monotonously decreases with increasing $\Omega_q$.
We only focus on the overall tendency in $\kappa$, neglecting the minor modulation in the data.
$\widetilde{V}_{\textup{JJ}}$ is the estimated mean ac-voltage bias amplitude across the JJ of the transmon. 
Even the maximum $\widetilde{V}_{\textup{JJ}}$ in the experimental range is less than the twice of the superconducting gap voltage of aluminium ($2\Delta_{\textup{Al}}/e$). 
Thus, we neglect the microscopic effects of Cooper-pair breaking directly induced by the resonator field.
The gray area represents 10\% scaling errors in $\Omega_q$ assuming imperfections in calibrating $P_r$.
See \textbf{Supplementary Notes 7} for details on $P_r$ and $\widetilde{V}_{\textup{JJ}}$.

\begin{figure}
    \centering
    \includegraphics[width=1.0\linewidth]{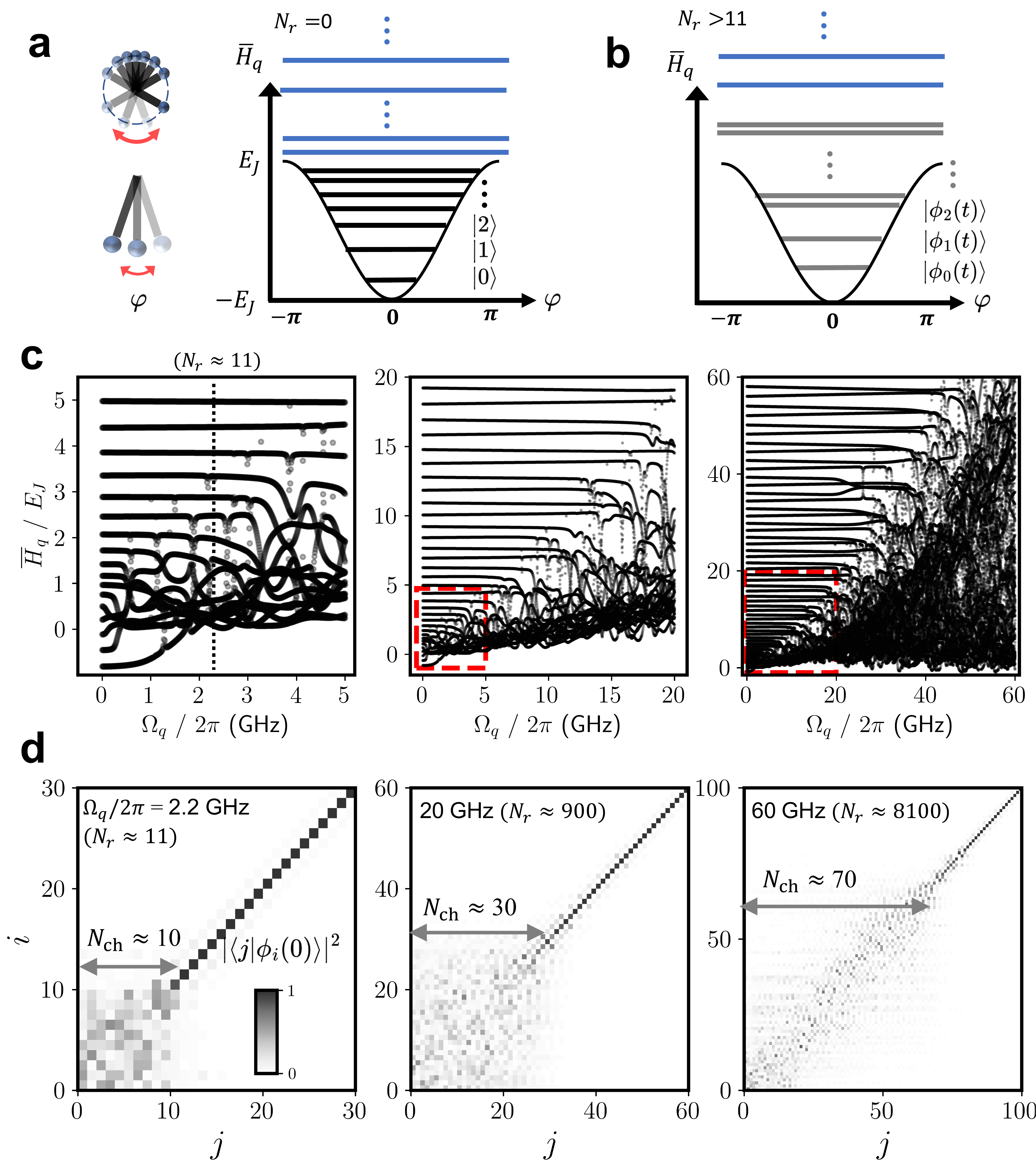}
    \caption{\textbf{Floquet modes of the transmon in the semi-classical approximation.} \textbf{(a–b)} Energy diagrams of the undriven and driven transmons, respectively. Each horizontal line indicates $\overline{H}_q$, the averaged energies of the eigenstates ${\ket{i}}$ or Floquet modes $\ket{\phi_{i}(t)}$. $\varphi$ and $E_J$ indicate superconducting phase and Josephson energy, respectively. In the strong drive limit, chaotic Floquet modes (gray lines) grow down to the ground state. The regular states (blue) lie above them. \textbf{(c)} Numerically calculated $\overline{H}_q$ with respect to $\Omega_q$. The dashed line indicate the threshold $\Omega_q$ and corresponding approximate $N_r$ above which the ground state falls into the chaotic layer.
    \textbf{(d)} Calculated $|\braket{j|\phi_i (0)}|^2$ for $\Omega_q/2\pi$=2.2, 20, and 60 GHz (from left to right). $N_{\textup{ch}}$ is the roughly estimated size of chaotic subspace, which linearly increases with $\Omega_q$. We use experimental parameters while setting $n_g=0.25$ and $\omega_d=\omega_r$ in the calculations.}
    \label{fig:2}
\end{figure}

Fig.~\ref{fig:1}(f) shows the resonator output spectrum at $N_r \approx 2000$ (blue) and background (gray). 
The background is fitted linearly (lower dashed), and the signal with a Lorentzian on top (upper dashed). 
The sharp spike corresponds to the transmitted probe ($S_{21}$), while the broad peak, with linewidth $\approx \kappa/2\pi$, is attributed to $\delta\hat{a}$.
The enclosed area corresponds to $\hbar\omega_r \kappa_{\mathrm{ex}}\langle\delta\hat{a}^\dagger\delta\hat{a}\rangle/2$. 
Fig.~\ref{fig:1}(g) plots $\langle\delta\hat{a}^\dagger\delta\hat{a}\rangle$ versus $N_r$. 
When $N_r\gtrsim400$, the linewidths of the sharp spike is nearly identical to that of the purely transmitted signal, confirming that inhomogeneous broadening in $S_{21}$ is negligible.
Since $N_r \gg \left<\delta\hat{a}^\dagger\delta\hat{a}\right>$ holds across the experimental range, the semiclassical approximation remains valid.
Experimentally exploring $\langle\delta\hat{a}^\dagger\delta\hat{a}\rangle$
with larger $N_r$ is inaccessible due to increasing instrumental noise of the probe becomes disturbing.
See \textbf{Supplementary Note 7} for supporting data and descriptions.

In Fig.~\ref{fig:2}, we numerically examine the Floquet modes of the transmon associated with $\Omega_q$.
Fig.~\ref{fig:2}(a-b) respectively illustrates the averaged energies $\overline{H}_q$ of the undriven and driven transmon ($\textbf{Methods}$).
Pendulum analogies for stable and unstable eigenstates (black and blue) are given beside.
Above the threshold, the Floquet modes ($\ket{\phi_i(t)}$) are classified into two classes, chaotic (gray, $i \leq N_{\textup{ch}}$) and regular modes (blue, $i > N_{\textup{ch}}$). Here, $N_{\textup{ch}}$ denotes the number of chaotic Floquet modes.

The numerical calculations of the Floquet modes obviously indicate that the unconfined states dominate the transmon dynamics in the experimental regime.
Fig.~\ref{fig:2}(c) shows numerically calculated $\overline{H}_q$ with $n_g$=0.25, $\omega_d=\omega_r$ and the experimental parameters. For $N_r\approx11$, where the bare resonator response appears in Fig.~\ref{fig:1}(c), the ground state falls into chaotic layer.
Such agreement is also expected from the previous work \cite{driven-system-20}.

Fig.~\ref{fig:2}(d) visualizes $|\braket{j|\phi_i(0)}|^2$ for $\Omega_q/2\pi$ = 2.2, 20, and 60 GHz (from left to right). $N_{\textup{ch}}$ indicates the number of the bases associated in the chaotic subspace, which increases linearly with increasing $\Omega_q$. We confirm $\ket{\phi_i(t)}$ involved in chaotic layer are the random superpositions of ${\ket{i}}$ for $0\leq i\leq N_{\textup{ch}}$, whereas the regular modes are proportional to ${\ket{i}}$.
We also perform the calculation for other $n_g$, and obtain the approximately same tendencies for $\overline{H}_q$ and $|\braket{j|\phi_i(0)}|^2$. 
See $\textbf{Methods}$ for summaries on the Floquet formalism and the numerical methods.
The Floquet-related numerical calculations throughout this work is based on QuTiP \cite{Qutip1,Qutip2}.

\section*{Energy loss at the transmon}
This section presents relevance between the observed ND and the energy loss rates from the transmon to its baths.
None of the existing ND models for a superconducting resonator capture the observed ND presented in Fig.~\ref{fig:1}(e–f). 
Nonlinear dissipation in superconducting resonators typically arises due to the interactions with two-level system baths or kinetic inductance. However, the scale and tendency of the observed ND in Fig.~\ref{fig:1}(e–f) are not consistent with those mechanisms.
Moreover, both mechanisms are known to contribute negligibly to ND over the explored range of $N_r$.
All these naturally suggest that dissipation induced by the transmon is responsible for the observed ND.

Let us define $\kappa_o$ the resonator dissipation irrelevant with the transmon, independent of $N_r$.
We can interpret $\hbar\omega_d N_r(\kappa - \kappa_o)$ as the nonlinear energy dissipation rate induced by the transmon.
We invoke two factors contributing to $\hbar\omega_d N_r(\kappa - \kappa_o)$.
The first is the energy loss rate occurring at the transmon in the semi-classical approximation ($\mathcal{T}$).
The other is the energy flow from the transmon into the resonator ($\mathcal{R}$), which yields the noise power spectrum shown in Fig.~\ref{fig:1}(f).
$\mathcal{R}$ should be a function of $\left<\delta\hat{a}^\dagger\delta\hat{a} \right>$.
Neglecting any correlated effects between $\mathcal{T}$ and $\mathcal{R}$, we have
\begin{equation}
\begin{split}
\label{eq3}
    \hbar\omega_d N_r(\kappa - \kappa_o)\approx \mathcal{T} + \mathcal{R},
\end{split}
\end{equation}
in steady states.

\begin{figure}
    \centering
    \includegraphics[width=0.9\linewidth]{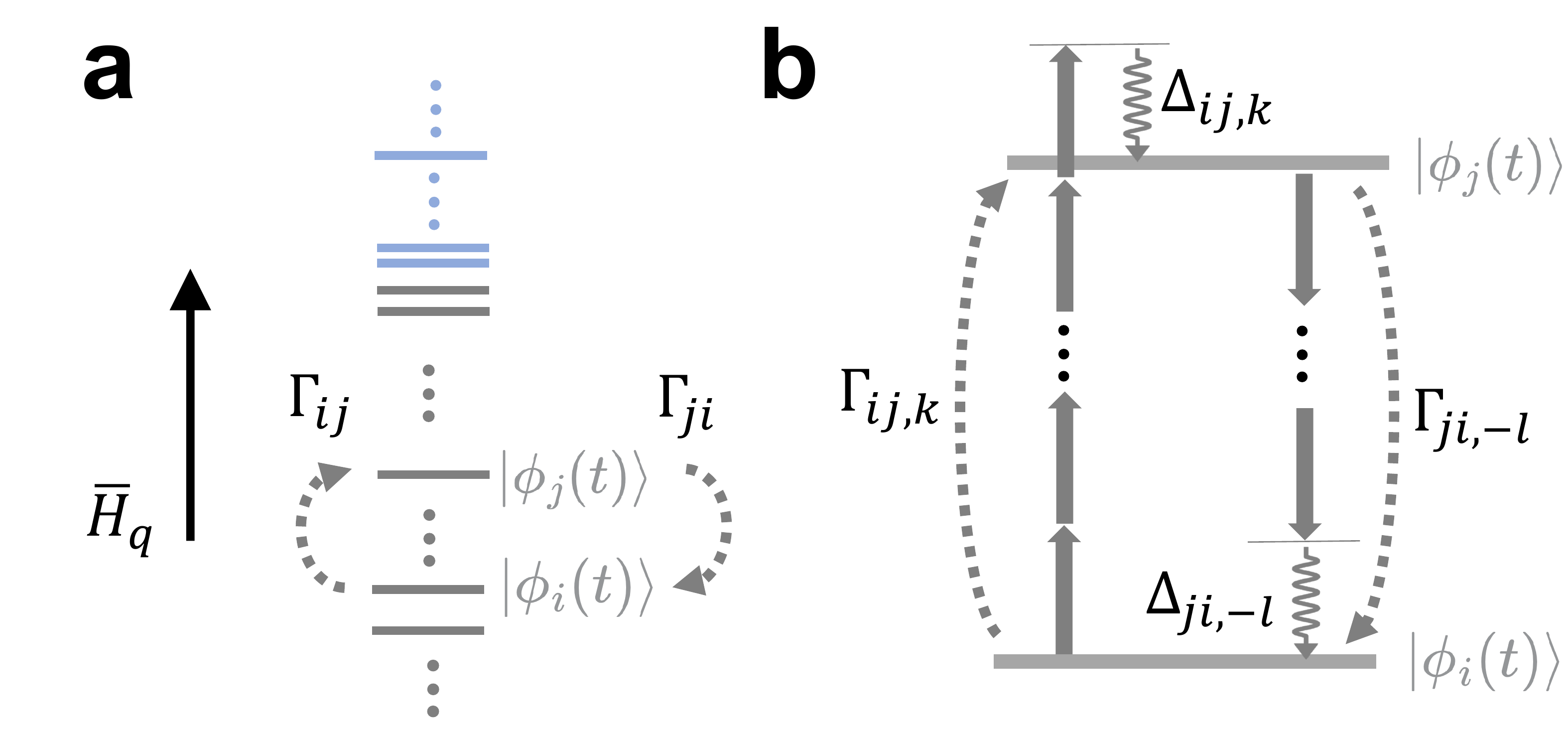}
    \caption{\textbf{Theoretical description for drive-induced energy loss at the transmon in the semiclassical approximation ($\mathcal{T}$).} \textbf{(a)} Transitions between two Floquet modes ($\ket{\phi_i(t)}$ and $\ket{\phi_j(t)}$) of the driven transmon in the semi-classical approximation. $\Gamma_{ij}$ and $\Gamma_{ji}$ indicates each transition rate. In the steady state, the relation $\Sigma_{i}p_{i}\Gamma_{ij} = \Sigma_{i}p_{i}\Gamma_{ji}$ is satisfied, where $p_i$ is the probability to find the system in $\ket{\phi_i(t)}$. \textbf{(b)} The resonator photons assist the transition between $\ket{\phi_i(t)}$ and $\ket{\phi_j(t)}$. The upward dashed arrows indicates a process assisted by $k$-photon absorption from the resonator photons (multiple upward solid arrows) at a rate of $\Gamma_{ij,k}$. The downward dashed arrows indicates a process assisted by $l$-photon emission into the resonator (multiple downward solid arrows) at a rate of $\Gamma_{ji,-l}$.
    The multiple solid arrows represent the resonator photons participating in the processes. $\Gamma_{ij}$ is then given by $\sum_{k}\Gamma_{ij,k}$.
    For each scenarios, the transmon emits a photon of energy  $\Delta_{ij,k} = \epsilon_i - \epsilon_j + k\hbar\omega_d$ ($\Delta_{ji,-l} = \epsilon_j - \epsilon_i - l\hbar\omega_d$) into its environments (wavy arrows). Here, $\epsilon_i$ is the quasienergies in the first Brillouin zone. Eventually, $\mathcal{T}$ is given by $\hbar{\sum_{ijk}\left( p_{i}{\Gamma_{ij,k} \Delta_{ij,k}}\right)}$.}
    \label{fig:3}
\end{figure}

Fig.~\ref{fig:3}(a) depicts the transitions between the Floquet modes of the driven transmon in the semi-classical approximation. 
$\Gamma_{ij}$ denotes the transition rate from $\ket{\phi_i(t)}$ to $\ket{\phi_j(t)}$.
In the steady state, the relation $\Sigma_{i}p_{i}\Gamma_{ij} = \Sigma_{i}p_{i}\Gamma_{ji}$ holds, where $p_i$ is the probability to find the system in $\ket{\phi_i(t)}$. 
Fig.~\ref{fig:3}(b) shows how the resonator photons contribute to the transition between $\ket{\phi_i(t)}$ and $\ket{\phi_j(t)}$. Two of all possible scenarios are presented. 
$\Gamma_{ij,k}$ ($\Gamma_{ji,-l}$) is the transition rate with $k$-photon absorption from the resonator ($l$-photon emission into the resonator). 
$\Gamma_{ij}$ is then given by $\sum_{k}\Gamma_{ij,k}$.
Eventually we have
\begin{equation}
\begin{split}
\label{eq4}
    {\mathcal{T}} \approx \hbar{\sum_{ijk}\left( p_{i}{\Gamma_{ij,k} \Delta_{ij,k}}\right)}.
\end{split}
\end{equation}
Here, $\Delta_{ij,k} = \epsilon_i - \epsilon_j + k\hbar\omega_d$ and $\epsilon_i$ are the quasienergies in the first Brillouin zone.
We hypothetically estimate $\mathcal{R}$ based on the observed  
$\left<\delta\hat{a}^\dagger\delta\hat{a} \right>$ in Fig.~\ref{fig:1}(g–h).

%

\section*{Numerical calculations}

\begin{figure*}
    \centering
    \includegraphics[width=0.8\linewidth]{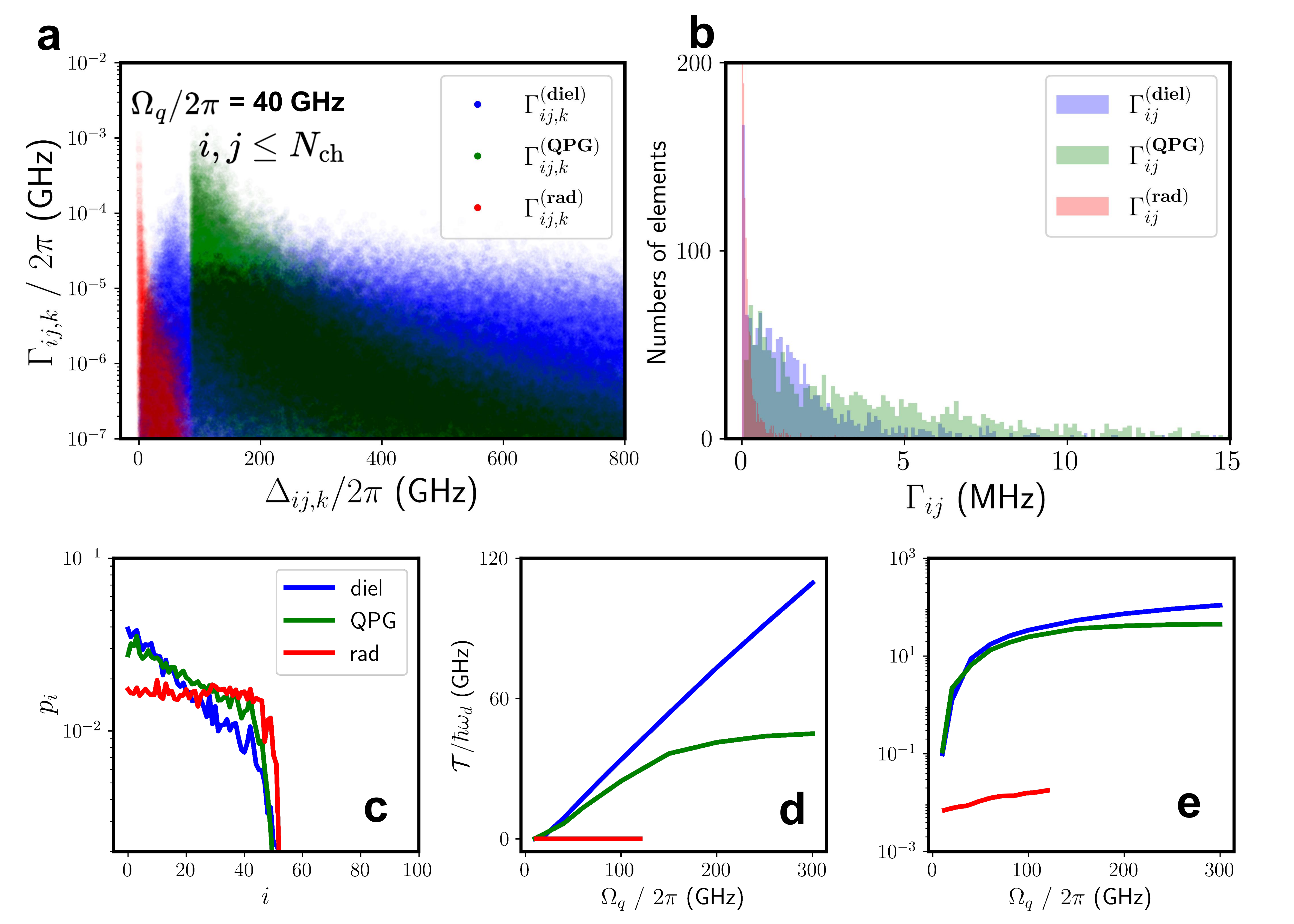}
    \caption{\textbf{Numerical calculations of drive-induced energy loss at the transmon in the semiclassical approximation ($\mathcal{T}$).} The Markovinity of the baths is assumed. Radiative (\textbf{rad}), dielectric (\textbf{diel}) and drive-induced quasiparticle generation (\textbf{QPG}) mechanisms are taken into consideration. We use $n_g=0.25$ and experimental values in $\hat{H}_q$ for the calculations. $J^{(\textbf{rad})}(\omega)$ is the Ohmic upper bound. In $J^{(\textbf{diel/QPG})}(\omega)$, we use $\omega_{\textbf{diel}}^c/2\pi=1$ THz and $\omega_{\textbf{QPG}}^c/2\pi=17$ GHz.
    \textbf{(a–c)} The drive amplitude $\Omega_q$ in the calculations is $2\pi \times$ 40 GHz. 
    \textbf{(a)} Calculated $\Gamma_{ij,k}$ for each dissipation mechanisms with respect to $\Delta_{ij,k}$.
    Only the elements with $i,j\leq N_{\textup{ch}}$ that dominantly affect the transmon dynamics are presented. \textbf{(b)} Histogram of calculated transition rates between Floquet modes $\ket{\phi_i(t)}$ and $\ket{\phi_j(t)}$ for each dissipation mechanisms. \textbf{(c)} Calculated  steady state population $p_i$ when $\Omega_q/2\pi=40$ GHz. \textbf{(d–e)} Calculated $\mathcal{T}\approx\hbar\omega_d N_r(\kappa - \kappa_o)$ when $\Gamma_{ij,k}$ is $\Gamma_{ij,k}^{(\textbf{rad})}$,$\Gamma_{ij,k}^{(\textbf{diel})}$, and $\Gamma_{ij,k}^{(\textbf{QPG})}$, respectively.
    The radiative calculations stop at $\Omega_q/2\pi=120$ GHz for a technical reason (See \textbf{Supplementary Note 4}).}
    \label{fig:4}
\end{figure*}
This section presents numerical calculations of $\mathcal{T}$. We introduce three major dissipation mechanisms: radiation, dielectrics, and drive-induced QPG \cite{QPG-1, QPG-2}.
They are labeled as $\textbf{rad}$, $\textbf{diel}$, and $\textbf{QPG}$, hereinafter.
Although our primary focus is QPG, we also include radiation and dielectric losses into our study for completeness.
Throughout this section, we assume the Markovianity of the baths.

We use the Floquet–Markov master equation \cite{driven-system-18, driven-system-20, FM1, FM2} to calculate $\Gamma_{ij,k}$ and $p_{i}$ in Eq.~\ref{eq4}.
$\Gamma_{ij,k}$ and $\Gamma_{ij}$ are given by
$\sum_{\textup{$\Psi$}}\Gamma_{ij,k}^{\textup{($\Psi$)}}$ and 
$\sum_{\textup{$\Psi$}}\Gamma_{ij}^{\textup{($\Psi$)}}$ ($\Psi$=$\textbf{rad}$, $\textbf{diel}$, and $\textbf{QPG}$).
Here, $\Gamma_{ij,k}^{\textup{($\Psi$)}}$ and $\Gamma_{ij}^{\textup{($\Psi$)}}$ are the transition rates caused by each mechanism.
$\textbf{Methods}$ summarizes the associated operators, normalized bath spectra (NBS) $J^{(\Psi)}(\omega)$ and their relation to $\Gamma_{ij,k}^{\textup{($\Psi$)}}$. 
We also introduce a QPG conductance $\sigma(\omega)$ by $J^{(\textbf{QPG})}(\omega)=\frac{\omega}{\pi g_K}\sigma(\omega)$. Here, $g_K$ is the inverse of Klitzing constant.
Fig.~\ref{fig:4}(a–b) present the calculated $\Gamma_{ij,k}^{\textup{($\Psi$)}}$ and $\Gamma_{ij}^{\textup{($\Psi$)}}$, respectively. 
Fig.~\ref{fig:4}(c–e) present the calculated $p_i$ and $\mathcal{T}$ for corresponding mechanisms.
The non-radiative mechanisms ($\textbf{diel}$ and $\textbf{QPG}$) dominate $\Gamma_{ij,k}$ and $\mathcal{T}$.

We use $J^{(\textbf{rad})}(\omega)=\omega/Q_{\textbf{rad}}$ \cite{Zhang-PRX-21} in Fig.~\ref{fig:4} without HFC. We use $Q_{\textbf{rad}}=3830$ in the calculation, which is the approximate lower bound.
The results indicate that radiative mechanism contributes negligibly to $\mathcal{T}$.
%
For the dielectric loss, we use  $J^{(\textbf{diel})}(\omega)=\frac{\hbar\omega^2}{4E_C Q_{\textbf{diel}}(\omega)}$ \cite{Zhang-PRX-21,Ye-SA-24}. $1/Q_{\textbf{diel}}(\omega)$ is the dielectric loss tangent of the transmon.
We hypothetically set $1/Q_{\textbf{diel}}(\omega)= (1/Q_{\textbf{diel}})e^{-\omega/\omega_{\textbf{diel}}^c}$, where $1/Q_{\textbf{diel}}$ is the effective constant loss tangent and $\omega_{\textbf{diel}}^c$ is the HFC of the dielectric bath spectrum.
We set $Q_{\textbf{diel}}=4.8\times10^5$, the realistic value for a transmon with decent device quality \cite{Ye-SA-24}.
We also set $\omega_{\textbf{diel}}^c/2\pi=1$ THz in Fig.~\ref{fig:4}.

Previous study~\cite{Houzet-PRL-2019} derives $J^{\mathrm{(\textbf{QPG})}}(\omega) = 0$ for $\omega < 2\Delta_{\textup{Al}}/\hbar$ and
$J^{\mathrm{(\textbf{QPG})}}(\omega) \approx \frac{8E_J}{\pi\Delta_{\textup{Al}}}\omega$ for $\omega \geq 2\Delta_{\textup{Al}}/\hbar$.
We call this ideal QPG model from now on.
The required experimental inputs are $E_J$ and aluminum gap energy $2\Delta_{\textup{Al}}$, which are known from the device parameter estimation and material property ($E_J/h=14.24$ GHz, and $\Delta_{\textup{Al}}=180\mu$eV).
Unfortunately, calculated $\mathcal{T}$ based on $J^{\mathrm{(\textbf{QPG})}}(\omega)$ with these values far exceeds the observations. The calculations hardly converge as increasing the range of $\omega$, and thus, we cannot even specify the $\mathcal{T}$ with our computational power at hand.
To tackle this problem, we introduce an effective inductance to the QPG conductance, which readily yields a HFC factor ${\left(1 + \left({\omega/\omega^c_{\textbf{QPG}}}\right)^2\right)^{-1}}$ to $J^{(\textbf{QPG})}(\omega)$ (\textbf{Supplementary Note 3}).
We set $\omega_{\textbf{QPG}}^c/2\pi=17$ GHz in Fig.~\ref{fig:4}, with which the calculated $\mathcal{T}$ becomes compatible to the experimental data.
Meanwhile, we neglect non-equilibrium QPs in the calculations (\textbf{Supplementary Note 3}).

The QPG event switches the charge parity of the transmon at the rates of $\sum_{ij}p_i\Gamma_{ij}^{(\textbf{QPG})}$.
Then, $n_g$ in $\hat{H}_q$ rapidly switches between $n_g^{s} - 0.25$ and $n_g^{s} + 0.25$, when the static charge offset is $n_g^{s}$.
Thus, the experimental results indicate the averaged effects of each parity.
We confirm that the calculated $\mathcal{T}$ with other $n_g$ are not exactly but approximately the same to those with $n_g=0.25$, and thus, computation with a single value of $n_g$ is sufficiently accurate.
\begin{figure}
    \centering
    \includegraphics[width=1.0\linewidth]{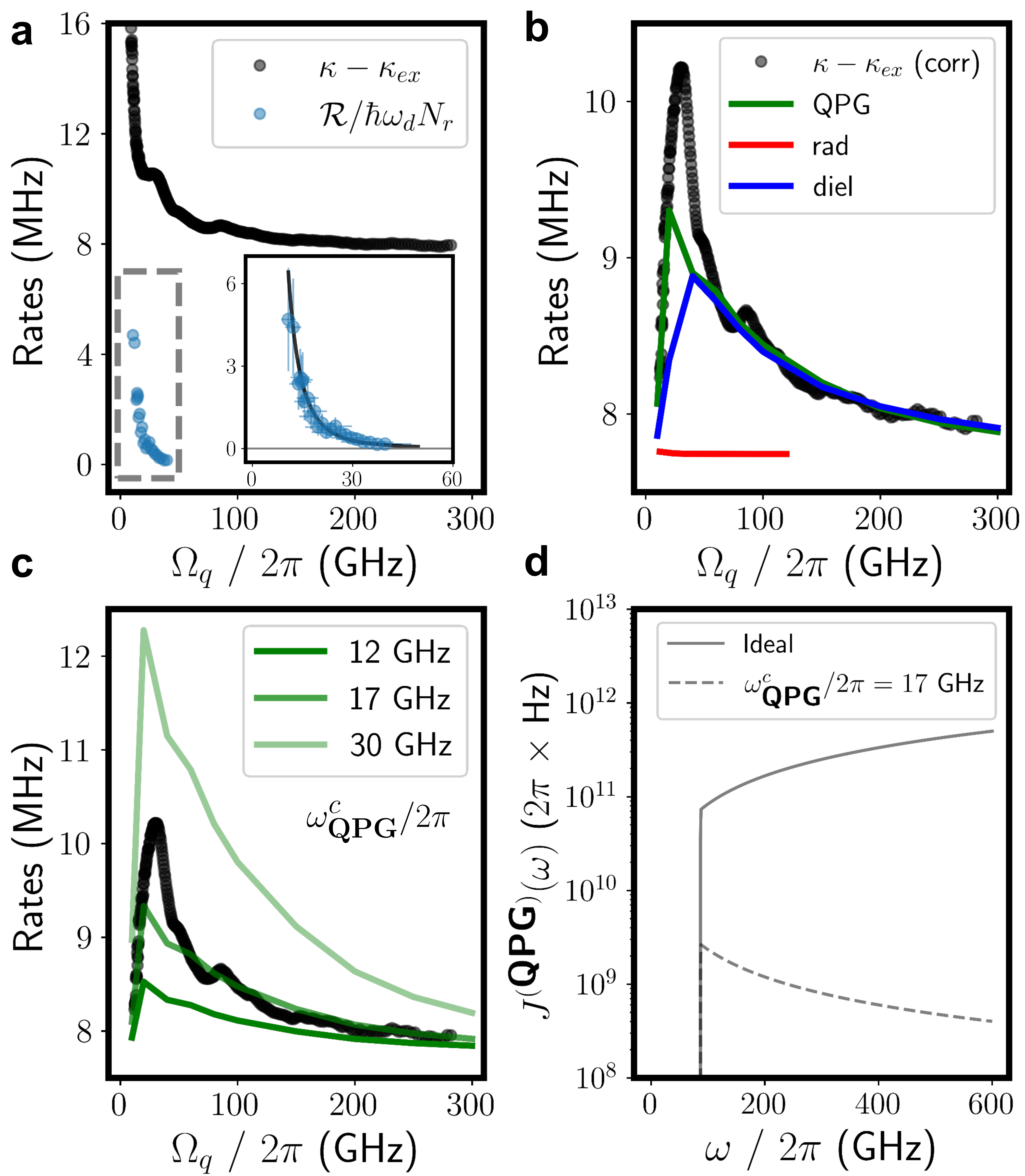}
    \caption{\textbf{Comparing calculations and experimental data.} 
    The Markovinity of the baths is assumed in the calculations. \textbf{(a)} Circles indicate the observed $\kappa - \kappa_{ex}$ (dark) and $\mathcal{R}/\hbar\omega_d N_r$ (bright blue). Hypothetically, we assume $\mathcal{R}=\hbar\omega_d\langle\delta\hat{a}^\dagger\delta\hat{a}\rangle\kappa$. The inset magnifies the data enclosed by the dashed box with error bars and a power fit. Error bars represent scaling errors assuming $\pm$20\% imperfect calibration in $\left<\delta\hat{a}^\dagger\delta\hat{a} \right>$ and $N_r$.
    From this, we invoke $\hbar\omega_d N_r(\kappa - \kappa_o)\approx \mathcal{T}$ for $\Omega_q/2\pi\gtrsim40$ GHz.    
    \textbf{(b)} Corrected $\kappa - \kappa_{ex}$ (circles) after subtracting $\mathcal{R}/\hbar\omega_d N_r$ from $\kappa$. We plot numerically calculated $ \mathcal{T}/\hbar\omega_d N_r + (\kappa_o - \kappa_{ex})$ based on each dissipation mechanism (lines). For non-radiative mechanisms, the used bath parameters (see the text) yield consistency with the experimental data of $\Omega_q/2\pi\gtrsim40$ GHz.
    Discrepancies in $\Omega_q/2\pi<40$ GHz persist and possibly result from the breakdown of $\mathcal{R}=\hbar\omega_d\langle\delta\hat{a}^\dagger\delta\hat{a}\rangle\kappa$ or unknown correlations between $\mathcal{T}$ and $\mathcal{R}$ not captured in Eq.~\ref{eq3}. 
    \textbf{(c)} Comparison between the experimental data and the calculations when taking only QPG into account. We only focus on the regime of $\Omega_q/2\pi\gtrsim40$ GHz, where we expect $\hbar\omega_d N_r(\kappa - \kappa_o)\approx \mathcal{T}$ is satisfied. 
    The opacity represents different $\omega^c_{\textbf{QPG}}/2\pi$ while $\kappa_o/2\pi$ is fixed by 16.82 MHz. We need a lower $\omega^c_{\textbf{QPG}}$ if the loss mechanisms other than QPG are significant. Thus, we can interpret 17 GHz as the approximate upper bound of $\omega^c_{\textbf{QPG}}/2\pi$. \textbf{(d)} Calculated $J^{(\textbf{QPG})}(\omega)$. We compare an ideal $J^{(\textbf{QPG})}(\omega)$ and that with $\omega^c_{\textbf{QPG}}/2\pi =17$ GHz.}
    \label{fig:5}
\end{figure}

The large non-radiative contribution to $\mathcal{T}$ originates from transitions among the unconfined states.
For the operators relevant for radiative ($\hat{n}$) and non-radiative dissipation ($\hat{\varphi}$, $\sin{\frac{\hat{\varphi}}{2}}$), their matrix representation in the confined subspace ($\ket{i}, i<7$) is similar, but dramatically different in the unconfined subspace ($i\geq7$).
Therefore, the substantial differences between the radiative and non-radiative $\Gamma_{ij,k}$ come from the dissipation of the unconfined states.

\section*{Comparing calculations and experimental data}

Fig.~\ref{fig:5}(a) shows the observed $\kappa-\kappa_{ex}$ (dark) and $\mathcal{R}/\hbar\omega_d N_r$ (bright blue).
Hypothetically, we assume $\mathcal{R}=\hbar\omega_d\kappa\langle\delta\hat{a}^\dagger\delta\hat{a}\rangle$ in Fig.~\ref{fig:5}(a).
The power fit (solid line in the inset) indicates that $\mathcal{R}/\hbar\omega_d N_r$ almost diminishes for $\Omega_q/2\pi\gtrsim40$ GHz.
Although the exact relation between $\mathcal{R}$ and $\langle\delta\hat{a}^\dagger\delta\hat{a}\rangle$ remains intractable, 
we expect that $\mathcal{R} = \hbar\omega_d\kappa\langle\delta\hat{a}^\dagger\delta\hat{a}\rangle$ holds at least to order-of-magnitude accuracy.
Consequently, $\hbar\omega_d N_r(\kappa - \kappa_o)\approx \mathcal{T}$ holds for $\Omega_q/2\pi\gtrsim40~\text{GHz}$.
It leads to a conclusion that $\delta\hat{a}$ contributes at most perturbatively to the transmon dynamics for $\Omega_q/2\pi\gtrsim40~\text{GHz}$.
Accordingly, using the semiclassical approximation in Eq.~\ref{eq4} is surely justified for $\Omega_q/2\pi\gtrsim40$ GHz.

Fig.~\ref{fig:5}(b) shows the corrected $\kappa$ (dark), obtained by subtracting $\mathcal{R}/\hbar\omega_d N_r$ from the measured $\kappa$. 
We compare it with the theoretically calculated $\mathcal{T}/\hbar\omega_d N_r + (\kappa_o - \kappa_{ex})$ based on Eq.~\ref{eq4}.
The red curve represents the radiative upper bound, which is negligible to the observation.
The blue curves correspond to the calculation based on the dielectric.
We find $J^{(\textbf{diel})}(\omega)$ with $Q_{\textbf{diel}}=4.8\times10^5$ and $\omega_{\textbf{diel}}^c/2\pi=1$ THz nicely fits the experimental data for $\Omega_q/2\pi\gtrsim40$ GHz.
Nonetheless, the form of $J^{(\textbf{diel})}(\omega)$ used in our calculation is hypothetical, and therefore the obtained bath parameters should not be interpreted too strictly.

The green curves correspond to the calculation based on the QPG mechanism.
Fixing $E_J/h=14.24$ GHz, and $\Delta_{\textup{Al}}=180\mu$eV, we find $\omega^c_{\textbf{QPG}}/2\pi=17$ GHz consistently fits the experimental data for $\Omega_q/2\pi\gtrsim40$ GHz.
Since $E_J$ and $\Delta_{\textup{Al}}$ are bounded from the device parameter and widely known material property, $\omega^c_{\textbf{QPG}}$ is the unique variable to determine.
Consequently, the fitted $\omega^c_{\textbf{QPG}}$ from the data can possess stringent physical meaning.
If the loss mechanisms other than QPG are significant, a lower value of 
$\omega^{c}_{\textbf{QPG}}$ is required.
The roughly estimated confident intervals of $\omega_{\textbf{QPG}}^c$ is approximately $\pm2\%$.

In Fig.~\ref{fig:5}(c), we further resolve $\omega_{\textbf{QPG}}^c$ from the data $\Omega_q/2\pi\gtrsim 40$ GHz. One can confirm that the calculations deviate from the experimental data with some variations in $\omega_{\textbf{QPG}}^c/2\pi$ from 17 GHz.
Consequently, we may interpret 17~GHz as an 
approximate upper bound for $\omega^{c}_{\textbf{QPG}}/2\pi$.
Fig.~\ref{fig:5}(d) compares the $J^{(\textbf{QPG})}(\omega)$ without cutoff (solid, ideal) and with $\omega^c_{\textbf{QPG}}/2\pi=17$ GHz (dashed). 
The cutoff reduces $J^{(\textbf{QPG})}(\omega)$ by approximately a few orders of magnitude.

We attribute the discrepancies between the experimental data and fits in $\Omega_q/2\pi<40$ GHz to the probable breakdown of $\mathcal{R}=\hbar\omega_d\langle\delta\hat{a}^\dagger\delta\hat{a}\rangle\kappa$ or unknown correlations between $\mathcal{T}$ and $\mathcal{R}$ not captured in Eq.~\ref{eq3}. 
Some minor discrepancies exist even in $\Omega_q/2\pi\gtrsim 40$ GHz. 
We attribute these to perturbative effects induced by $\delta\hat{a}$ not captured in the semiclassical approximation.

We perform the same measurements and analysis with multiple devices having different parameters, demonstrating the robustness of our claims (\textbf{Extended Data}). 
Since our JJs were fabricated using standard procedures, we expect similar behavior to occur in other SCQ devices produced in the same way.

\section*{Discussions}
\subsection*{Interpretation and implications}

Our data indicate a breakdown of either the ideal QPG conductance model or the Markovian approximation applied to QPG processes.
Assuming that the Markovianity of the QPG bath remains valid, we find that introducing a high-frequency cutoff (HFC) at $\omega^c_{\textbf{QPG}}/2\pi = 17-20~\mathrm{GHz}$ in the QPG bath spectrum reconciles the discrepancy between theoretical calculations and experimental observations.
These HFCs suppress $J^{(\textbf{QPG})}(\omega)$ by several orders of magnitude compared to the ideal model.
Alternatively, the discrepancy may signal a breakdown of the Markovian approximation itself.
Non-Markovianity arises when the bath memory time is not negligibly short compared with the characteristic timescale of the system dynamics.
In the present context, a finite bath memory time implies that the bath spectrum necessarily exhibits cutoffs or resonant structures.
Consequently, even in this case, the observed behavior effectively corresponds to a non-ideal form of \(J^{(\textbf{QPG})}(\omega)\).

If the non-ideal $J^{(\textbf{QPG})}(\omega)$ is indeed physical, it conveys an important practical implication for superconducting quantum computing. The expected next big challenge, drive-induced QPG, may be far less detrimental than previously expected.
Recognizing this provides correct guidance for operating superconducting quantum processors with minimal errors.

To the best of our knowledge, the findings in this study has not been explicitly addressed despite the long history of Josephson physics and decades of studies on dissipation in superconducting circuits.
In this regard, our study may point to an important open question in these fields.
In addition, our findings are crucial for gaining insight into both fundamentally or practically intriguing phenomena \cite{driven-system-16, driven-system-17, Fitzpatrick-PRX-17, Fink-PRX-17,Sett-PRXQ-24, Reed-PRL-10, Wang-NC-22, Majumder-JLTP-22, Nojiri-PRB-24} emerging in the macroscopic photon number regime of cQED.
Furthermore, deeper understanding on QPG can also contribute to designing the next-generation superconducting circuits less challenged by the structural instabilities of JJs \cite{Verney-PRAppl-19, Burgelman-PRAppl-22}, and properly comprehending the detrimental effects related with the unconfined state \cite{driven-system-19,driven-system-20,driven-system-21,driven-system-22,Majumder-JLTP-22,Nojiri-PRB-24}.

\subsection*{Other methods of measuring drive-induced QPG}

Several previous studies~\cite{Graaf-PRL-2013, Diamond-PRXQ-2023, Liu-PRL-2024} present experiments on QPG under high frequency radiations above the superconducting gap. Such experiments can possibly provide more direct clues on $J^{(\textbf{QPG})}(\omega)$.
Nonetheless, these studies focus on largely different directions rather than quantifying $J^{(\textbf{QPG})}(\omega)$. Furthermore, estimating the drives across the JJs considering all the unknown high frequency modes of the packaging have remained a challenge for accurate estimation in these papers. Such technical difficulty can be possibly solved in principle but is unavailable as of now.
See \textbf{Supplementary Note 8} for more discussion.

The drive-induced QPG rate can, in principle, be directly extracted from parity-switching measurements of transmons under sufficiently strong microwave drives. However, the drive amplitude cannot be made arbitrarily large, as excessive drive amplitude leads to drive-induced unwanted state transitions (DUST) and degrades the coherence properties of transmons, both of which hinder reliable parity-switching measurements.
For conventional drive frequencies in the 4–8 GHz range, the expected QPG rates are extremely small at moderate drive amplitudes (below 10 GHz). Distinguishing such sparse parity-switching events from those induced by the other mechanisms would be experimentally challenging.
Moreover, a quantitative assessment of QPG requires separating the contribution from drive-induced QP tunneling from that of non-equilibrium QP. Completely ruling out the contribution from the non-equilibrium QP needs information on their distribution. This demands additional experiments with extensive electromagnetic shielding as in \cite{Connolly-PRL-24}.

A possible alternative route is to employ substantially higher drive frequencies with accurate drive amplitude calibrations. Then, one can generate significantly larger QPG rates even with moderate drive amplitudes. This allows us to perform quantitative analysis without suffering from the degradation of the readout performance.
The device presented in a recent study \cite{Mencia-25} meets these conditions. This approach will be a promising direction for future experimental investigation on QPG at driven transmons.

\section*{Methods}

\subsection*{Device fabrication}
The circuit consists of a transmon and co-planar waveguide resonators (CPWRs). 
Only the fundamental mode of one of the CPWRs is utilized and considered in the model. The base layer was patterned in a 100 nm niobium titanium nitride (NbTiN) film on a 525 $\mu$m thick high-resistance silicon substrate. The Josephson junctions of the transmons are an Al-AlOx-Al structure fabricated using double-angle shadow evaporation. The overlapped areas of the constituent Josephson junctions are approximately $100\times200~nm^2$. See \textbf{Supplementary Note 7} for detailed descriptions.

\subsection*{Theoretical methods}

\subsubsection*{Floquet formalism}

We define $\ket{\phi_i^{\alpha}(t)} = e^{i\alpha\omega_d t}\ket{\phi_i^{0}(t)}$ and $\epsilon_{i}^{\alpha}=\epsilon_{i}^0+\alpha\hbar\omega_d$ ($\alpha \in \mathbb{Z}$) as Floquet modes and corresponding quasi-eigenenergies of $\hat{H}_q(t)$ with an Floquet order of $\alpha$.
Here, $\ket{\phi_i^{0}(t)}$ and $\epsilon_{i}^0$ are Floquet modes and corresponding quasi-eigenenergies in the first Brillouin zone.
We define $\ket{\phi_i(t)}=\ket{\phi_i^{0}(t)}$ and $\epsilon_{i}=\epsilon_{i}^0$ in the main text.
For a Floquet mode $\ket{\phi_i(t)}$, we define $\overline{H}_q$ 
\begin{equation}
\begin{split}
\label{eq_mm1}
    \overline{H}_q = \frac{\omega_d}{2\pi} \int_{0}^{2\pi/\omega_d} dt \bra{\phi_i(t)}\hat{H}_q(t)\ket{\phi_i(t)}.
\end{split}
\end{equation}
In the main text, we arrange Floquet modes such that $\ket{\phi_i(t)}$ yielding small $\overline{H}_q$ possess lower indexes $i$. 
Detail explanations are provided in \textbf{Supplementary Note 2}.

\subsubsection*{Floquet-Markov master equation}
We symbolize a specific transmon dissipation mechanism as `$\Psi$' when the associated dimensionless operators describing the transmon–bath interaction are $\hat{\Psi}^{(l)}$.
We label each operators with $l$. We introduce $J^{(l)}(\omega)$, the normalized bath spectra (NBS) of the corresponding operators.
Then, Floquet-Markov formalism yields
\begin{equation}
\begin{split}
\label{eq_mm2}
    \Gamma_{ij,k}^{(\Psi)} =\sum_{l}{J^{(l)}(\Delta_{ij,k})|\hat{\Psi}^{(l)}_{ij,k}|^2},
\end{split}
\end{equation}
where
\begin{equation}
\begin{split}
\label{eq_mm3}
   \hat{\Psi}_{ij,k}^{(l)} = \frac{\omega_d}{2\pi} \int_{0}^{2\pi/\omega_d} dt  \bra{\phi_i(t)}\hat{\Psi}^{(l)}\ket{\phi_j(t)}e^{-ik\omega_d t}.
\end{split}
\end{equation}
We take the zero-temperature limit ($\hbar|\Delta_{ij,k}|\gg k_B \Theta$, $k_B$ and $\Theta$ respectively refer to Boltzmann constant and the system temperature) in Eq~\ref{eq_mm2}.
Cooperating all the mechanisms into consideration, $\Gamma_{ij,k}$ and $\Gamma_{ij}$ are given by
$\sum_{\textup{$\Psi$}}\Gamma_{ij,k}^{\textup{($\Psi$)}}$ and 
$\sum_{\textup{$\Psi$}}\Gamma_{ij}^{\textup{($\Psi$)}}$, respectively.
More detailed descriptions on the above are provided in \textbf{Supplementary Note 2}.

Below, we write the explicit forms of $\Gamma_{ij,k}^{(\Psi)}$ for $\Psi=\textbf{rad},\textbf{diel}$, and $\textbf{QPG}$,
\begin{equation}
\begin{split}
\label{eq_mm4}
    \Gamma_{ij,k}^{(\textbf{rad})} ~ = &J^{\textbf{(rad)}}(\Delta_{ij,k})|\hat{n}_{ij,k}|^2,       \\
    \Gamma_{ij,k}^{(\textbf{diel})}~ = &J^{\textbf{(diel)}}(\Delta_{ij,k})|\hat{\varphi}_{ij,k}|^2,\\
    \Gamma_{ij,k}^{(\textbf{QPG})}   = &J^{(\textbf{QPG}+)}(\Delta_{ij,k})|\sin{(\hat{\varphi}/2)}_{ij,k}|^2 \\ + &J^{(\textbf{QPG}-)}(\Delta_{ij,k})|\cos{(\hat{\varphi}/2)}_{ij,k}|^2,
\end{split}
\end{equation}
where the normalized bath spectra (NBS) therein are given as
\begin{equation}
\begin{split}
\label{eq_mm5}
    &J^{\textbf{(rad)}}(\omega) =  \frac{\omega}{Q_{\textbf{rad}}},      \\
    &J^{\textbf{(diel)}}(\omega)= \frac{\hbar\omega^2}{4E_C Q_{\textbf{diel}}(\omega)},       \\
    &J^{(\textbf{QPG}+)}(\omega) = \frac{16E_J}{h}\frac{S^{+}(\omega)}{\left(1 + \left(\frac{\omega}{\omega^c_{\textbf{QPG}}}\right)^2\right)},       \\
    &J^{(\textbf{QPG}-)}(\omega) = \frac{16E_J}{h}\frac{S^{-}(\omega)}{\left(1 + \left(\frac{\omega}{\omega^c_{\textbf{QPG}}}\right)^2\right)}, 
\end{split}
\end{equation}
and
\begin{equation}
\begin{split}
\label{eq_mm6}
    &S^{\pm}(\omega) =  \int_{1}^{\infty} \int_{1}^{\infty}dx dy\frac{xy\pm1}{\sqrt{x^2-1}\sqrt{y^2-1}}\delta(\frac{\hbar\omega}{\Delta_{\textup{Al}}}-x-y).
\end{split}
\end{equation}

In $J^{(\textbf{QPG}\pm)}(\omega)$, we introduce cutoff factors with cutoff frequencies $\omega_{\textbf{QPG}}^c$.
Please also recall that we hypothetically set $1/Q_{\textbf{diel}}(\omega)= (1/Q_{\textbf{diel}})e^{-\omega/\omega_{\textbf{diel}}^c}$.
Here, $\delta(x)$ is a dirac-delta function and $\Delta_{\textup{Al}}$ is the superconducting gap energy of aluminum. 

$S^{\pm}(\omega)$ can be approximated as below
\begin{equation}
\begin{split}
\label{eq_mm7}
    S^{\pm}(\omega) =  0 ~~~~~~~~~~~~~~~~~~~~~~&(\frac{\hbar\omega}{\Delta_{\textup{Al}}} < 2), \\
    S^{+}(\omega) \approx  \pi[1+(\frac{\hbar\omega}{\Delta_{\textup{Al}}}+2)/4] ~~&(2<\frac{\hbar\omega}{\Delta_{\textup{Al}}} \ll 4),\\
    S^{-}(\omega) \approx  (\pi/2)(\frac{\hbar\omega}{\Delta_{\textup{Al}}}-2) ~~~~~~~&(2<\frac{\hbar\omega}{\Delta_{\textup{Al}}} \ll 4),\\
    S^{\pm}(\omega) \approx  \frac{\hbar\omega}{\Delta_{\textup{Al}}} ~~~~~~~~~~~~~~~~~~~~~~&(\frac{\hbar\omega}{\Delta_{\textup{Al}}} \gg 4).
\end{split}
\end{equation}
For calculating matrix elements $|\hat{n}_{ij,k}|$, $|\hat{\varphi}_{ij,k}|$, $|\sin{(\hat{\varphi}/2)}_{ij,k}|$, and $|\cos{(\hat{\varphi}/2)}_{ij,k}|$ in $\Gamma_{ij,k}^{({\Psi})}$, we use the following relations
\begin{equation}
\begin{split}
\label{eq_mm8}
    \bra{C_n}\hat{n}\ket{C_m} &= \delta_{nm}, \\
    \bra{C_n}\hat{\varphi}\ket{C_m} &= i\frac{(-1)^{(n-m+1)}}{n-m}~~(n \neq m), \\
                                    &= 0~~~~~~~~~~~~~~~~~~~(n = m), \\
    \bra{C_n}\sin{(\hat{\varphi}/2)}\ket{C_m}&= \frac{i}{\pi}\frac{(-1)^{(n-m+1)}(n-m)}{(n-m)^2-\frac{1}{4}},\\
    \bra{C_n}\cos{(\hat{\varphi}/2)}\ket{C_m}& = \frac{i}{2\pi}\frac{(-1)^{(n-m)}}{(n-m)^2-\frac{1}{4}},
\end{split}
\end{equation}
where $\delta_{nm}$ is a Kronecker-delta symbol and $\ket{C_m}$ is an eigenstate of $\hat{n}$ with an eigenvalue $m$ ($0,\pm1, \pm2, \cdots$). 
We introduce the imaginary unit $i$ in Eq.~\ref{eq_mm8}.
Supporting discussions on the above statements are provided in \textbf{Supplementary Note 3}.

From the numerical simulation, we confirm $|\cos{(\hat{\varphi}/2)}_{ij,k}|^2 \ll |\sin{(\hat{\varphi}/2)}_{ij,k}|^2$ is satisfied except when $i=j$ and $k=0$. 
However, these exceptional components never contribute to the $\Gamma_{ij,k}^{(\textbf{QPG})}$.
Therefore, $\Gamma_{ij,k}^{(\textbf{QPG})}$ should be limited by $J^{(\textbf{QPG}+)}(\omega)$.
In the main text and other parts of this work, we only consider $J^{(\textbf{QPG}+)}(\omega)$ to calculate $\Gamma_{ij,k}^{(\textbf{QPG})}$. We define $J^{(\textbf{QPG}+)}(\omega)$ and $\omega_{\textbf{QPG}+}^c$ as $J^{(\textbf{QPG})}(\omega)$ and $\omega_{\textbf{QPG}}^c$ in the main text, respectively.

\subsection*{Numerical methods}

Floquet-related numerical calculations in this work is based on QuTiP (version 4).
We utilize `\textsf{floquet$\_$modes}' function to compute $\ket{\phi_i^{0}(0)}$ and $\epsilon_{i}^0$. To compute their time-dependence over a period, we utilize `\textsf{floquet$\_$modes$\_$table}' function.
$\hat{\Psi}_{ij,k}^{(l)}$ and $\Gamma_{ij,k}$ are computed using `\textsf{floquet$\_$master$\_$equation$\_$matrix}' and `\textsf{floquet$\_$master$\_$equation$\_$rate}' functions, respectively. 
Once all  $\Gamma_{ij,k}$ are given, we compute steady state $p_i$ using `\textsf{floquet$\_$markov$\_$mesolver}'.
We compute each procedure separately to facilitate debugging and prevent memory saturation.

Detail explanations, tips, and conditions in the calculations are provided in \textbf{Supplementary Note 4}. 
Supporting data and explanations on the accuracy tests on our calculations are provided in \textbf{Supplementary Note 5}.
Hilbert space dimensions of 201 and 401 are used in calculating the radiative and non-radiative dissipations, respectively. See \textbf{Supplementary Note 4} for the rationale behind using a smaller dimension in the radiative case.
We also offer calculations with various bath parameters in \textbf{Supplementary Note 5}.

\subsection*{Experimental methods}
\subsubsection*{Cryogenic configuration}
The device is mounted on the mixing-chamber plate of a dilution refrigerator (Bluefors LD-400). 
Input and output signals are sent through separate lines, with 10/10/10/20 dB attenuators installed at the 4 K, 700 mK, 100 mK, and MXC stages of the input line, and a 40 dB isolator (LNF-ISISC) placed at the MXC stage of the output line.
A HEMT amplifier (LNF-LNC) is positioned at the 4 K stage of the output line. 
The device is enclosed in radiation and magnetic shieldings made of copper and cryoperm. Details are provided in \textbf{Supplementary Note 6}.

\subsubsection*{Measurement}
Transmission spectroscopy is performed using a two-port network analyzer (Keysight E5071C). 
We characterize the transmon with an integrated quantum control platform (Quantum Machines OPX-plus and Octave). 
For power spectrum measurements, we employ a signal generator (Keysight E8257D) and a spectrum analyzer (Signal Hound SA-124AB).
The output signals from the refrigerator are amplified with a 30 dB power amplifier (Narda MITEQ LNA-30-04000800) before being sent to the room-temperature instruments. 
Further details are provided in \textbf{Supplementary Note 6}.

\makeatletter
\renewcommand{\fnum@figure}{\textbf{Extended Data~}\thefigure}
\setcounter{figure}{0}
\makeatother

\begin{figure*}
    \centering
    \includegraphics[width=0.9\linewidth]{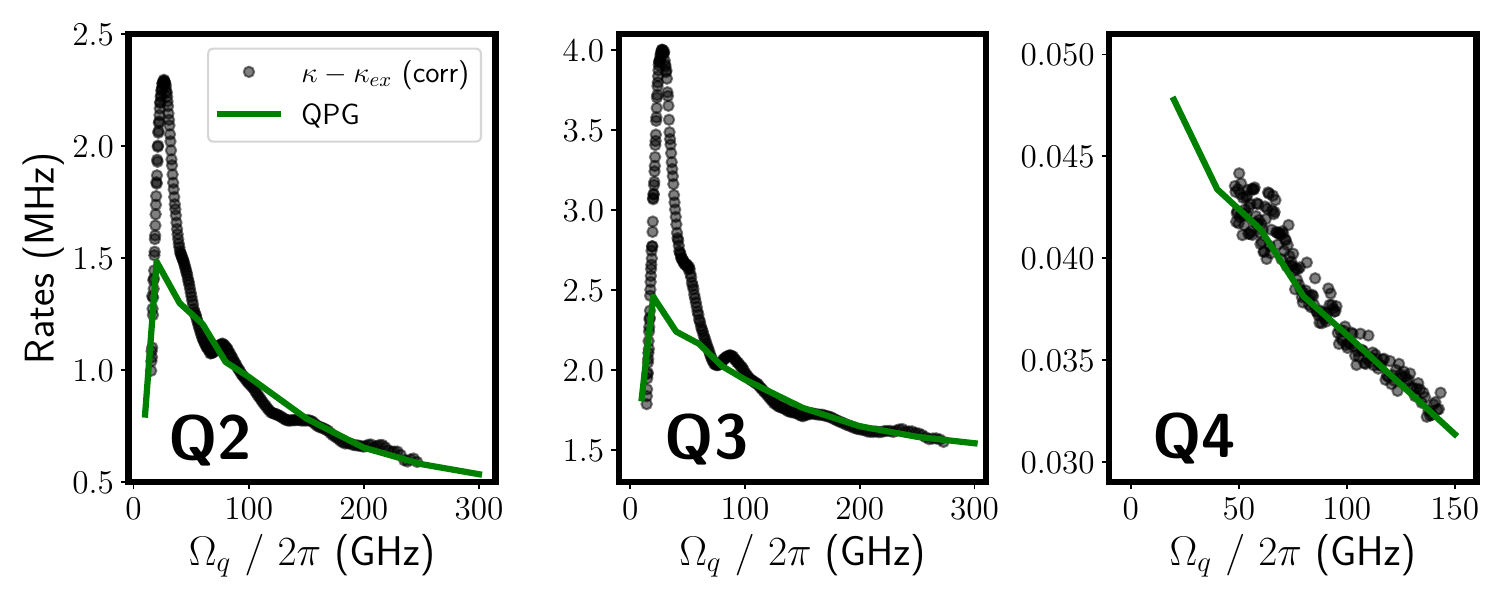}
    \caption{\textbf{Experimental data from Q2–Q4 and theoretical calculations based on QPG mechanism.} The circles indicate the corrected $\kappa$ subtracted by $\kappa_{ex}$ with respect to $\Omega_q/2\pi$ from the readout resonators of the other transmons (Q2–Q4).
    Here, the correction means subtracting $\mathcal{R}/\hbar\omega_d N_r$ from $\kappa$. The lines indicate the theoretically calculated $\mathcal{T}/\hbar\omega_d N_r + \kappa_o$ subtracted by $\kappa_{ex}$ when only taking QPG mechanism into consideration.
    As with Q1 in the main text, we assume that $\mathcal{T}$ mainly contribute to nonlinear dissipation in the regime $\Omega_q/2\pi \gtrsim 40$ GHz. We manually set $\omega_{\textbf{QPG}}^c/2\pi$ and $\kappa_{o}$ such that the theoretical curves consistently fit the experimental data over this regime. In the calculations, $\kappa_{o}$ is 16.79, 27.59, and 6.43 MHz, and $\omega_{\textbf{QPG}}^c/2\pi$ is 17, 19, and 20 GHz for Q2, Q3, and Q4, respectively. These values can be interpreted as approximate upper bounds of $\omega_{\textbf{QPG}}^c/2\pi$. We obtain clear and stable $S_{21}$ only above $\Omega_q/2\pi\approx50$ GHz for Q4, and thus, the range of $\Omega_q/2\pi\lesssim50$ GHz remains blank. 
    See \textbf{Supplementary Table 2} for the specifications of Q2–Q4 and their readout resonators. The comparably weak nonlinearity in $\kappa$ of Q4 arises from the significantly smaller transmon–resonator coupling compared to the others (see \textbf{Supplementary Note 8} for further explanation). Systematic and statistical errors are negligible in all the cases.}
    \label{fig:ext}
\end{figure*}

\section*{Extended data}
In this work, we investigated four transmons (Q1–Q4). Q1 is the one presented in the main text.
The others (Q2–Q4) possess the same circuit configuration but different system parameters. They yield the similar tendencies in $\kappa$ as for Q1.
We also obtain similar $\omega_{\textbf{QPG}}^c$ from Q2–Q4.
Extended Data \ref{fig:ext} presents the experimental results for Q2–Q4 together with theoretical calculations used to estimate the upper bounds of $\omega_{\textbf{QPG}}^c$. We assume that $\mathcal{T}$ dominates the nonlinear dissipation in the regime $\Omega_q/2\pi \gtrsim 40$ GHz for Q2–Q4 as in Q1.
The numerical simulation data already indirectly supports this assumption as showing consistencies to the observation in the regime $\Omega_q/2\pi \gtrsim 40$ GHz.

\section*{Acknowledgments}
We are grateful to Moonjoo Lee for fruitful discussions.
We thank David Thoen and Jochem Baselmans for providing us with NbTiN films.
This work was supported by the National Research Foundation of Korea (NRF) grant funded by the Korea government (MSIT)(RS-2023-00213037).
This work was also supported by Korea Research Institute of Standards and Science (KRISS-GP2025-0010-03 and KRISS-GP2025-0014-07) and the National Research Council of Science \& Technology(NST) grant by the Korea government (MSIT) (No. GTL25011-110).

\subsection*{Author contributions}
\textbf{B.A} (Conceptualization, Data curation, Formal analysis, Funding acquisition, Investigation, Methodology, Project administration, Resources, Software, Supervision, Validation, Visualization, Writing – original draft, Writing – review \& editing), 
\textbf{S.C} (Investigation, Validation, Writing – review \& editing), 
\textbf{H.C.P} (Validation, Writing – review \& editing), 
\textbf{S.D} (Resources, Writing – review \& editing), 
\textbf{R.D} (Resources, Writing – review \& editing), 
\textbf{G.A.S} (Resources, Writing – review \& editing), 
\textbf{J.K.} (Validation, Writing – review \& editing), 
\textbf{S.S} (Funding acquisition, Project administration, Resources, Writing – review \& editing), 
\textbf{J.S} (Validation, Funding acquisition, Project administration, Resources, Writing – review \& editing).

\subsection*{Competing interests}
The authors declare no competing interests.

\subsection*{Data and materials availability}
Data supporting the plots within the main text of this paper are available through Zenodo at http://10.5281/zenodo.17934039.
Code used to produce the plots and numerical data within this paper is available through Zenodo at http://10.5281/zenodo.17934460.

Further information is available from the corresponding author upon reasonable request.

\end{document}


\title{Supplementary Information for  \\ ``Evidence for unexpectedly low quasiparticle generation rates \\ across Josephson junctions of driven superconducting qubits''}
\author{Byoung-moo Ann}
\email{byoungmoo.ann@gmail.com}
\altaffiliation{Responsibility for the preparation of the supplementary materials lies solely with the author; the co-authors contributed through review.}
\affiliation{Quantum Technology Institute, Korea Research Institute of Standards and Science, 34113 Daejeon, South Korea}
\author{Sang-Jun Choi}
\affiliation{Department of Physics Education, Kongju National University, Gongju 32588, South Korea} 
\author{Hee Chul Park}
\affiliation{Department of Physics, Pukyong National University, Busan 48513, Korea} 
\author{Sercan Deve}
\affiliation{Kavli Institute of Nanoscience, Delft University of Technology, 2628 CJ Delft, The Netherlands} 
\author{Robin Dekker}
\affiliation{Kavli Institute of Nanoscience, Delft University of Technology, 2628 CJ Delft, The Netherlands} 
\author{Gary A. Steele}
\affiliation{Kavli Institute of Nanoscience, Delft University of Technology, 2628 CJ Delft, The Netherlands} 
\author{Jaseung Ku}
\affiliation{Quantum Technology Institute, Korea Research Institute of Standards and Science, 34113 Daejeon, South Korea}
\author{Seung-Bo Shim}
\affiliation{Quantum Technology Institute, Korea Research Institute of Standards and Science, 34113 Daejeon, South Korea}
\author{Junho Suh}
\affiliation{Quantum Technology Institute, Korea Research Institute of Standards and Science, 34113 Daejeon, South Korea}
\affiliation{Department of Physics, Pohang University of Science and Technology (POSTECH), Pohang,  South Korea}
\date{\today}

\maketitle

\tableofcontents

\clearpage

\makeatletter
\renewcommand{\fnum@figure}{Fig.~\thefigure}
\makeatother

\makeatletter
\renewcommand{\fnum@table}{Table~\thetable}
\makeatother
\newcommand{\hbAppendixPrefix}{S}
\renewcommand{\thefigure}{\hbAppendixPrefix\arabic{figure}}
\setcounter{figure}{0}
\renewcommand{\thetable}{\hbAppendixPrefix\arabic{table}} 
\setcounter{table}{0}
\renewcommand{\theequation}{\hbAppendixPrefix\arabic{equation}}
\setcounter{equation}{0}

\clearpage

\section*{Supplementary Note 1 : Model Hamiltonian}

\subsectionntoc{System simplification}
This section explains how our system can be reduced to Fig.1(a) in the main text.
Each circuit used in the experiments consists of a transmon and dispersively coupled co-planar waveguide resonators (CPWRs). 
See \textbf{Supplementary Note 8} for the detailed information on the circuits.
Let us define $\hat{H}_{\textup{full}}(t)$ as the full system Hamiltonian considering all the resonator modes. 
Then, $\hat{H}_{\textup{full}}(t)$ reads
\begin{equation}
\begin{split}
\label{eq_full}
    \hat{H}_{\textup{full}}(t)  =  4E_C(\hat{n} - n_g)^2 - E_J\cos\hat{\varphi}& +  \hbar\omega_r\hat{a}^\dagger\hat{a} - i\hbar g\hat{n}(\hat{a}- \hat{a}^\dagger) + \Omega_{r} (\hat{a}+\hat{a}^{\dagger}) \sin \omega_d t \\ & +  \sum_{k} \hbar\omega_{r,k}\hat{a}_{k}^\dagger\hat{a}_{k} - \sum_{k} i\hbar g_k\hat{n}(\hat{a}_{k}- \hat{a}_{k}^\dagger) + \Omega_{r,k} (\hat{a}_{k}+\hat{a}_{k}^{\dagger}) \sin \omega_d t.
\end{split}
\end{equation}
The definitions of $\hat{a}$, $\hat{n}$, and $\hat{\varphi}$, $E_C$, $E_J$, $g$, $\omega_d$, $\omega_r$, and $\Omega_r$ are the same as those in the main text.
We only use the fundamental mode of one of the CPWRs coupled to the transmon of each circuit, and its field operator is denoted by $\hat{a}$.
Meanwhile, $\hat{a}_{k}$ refer to a field operators of a resonator modes disregarded in the main text.
$\omega_{r,k}$ refers to the resonance frequencies of these modes. 
The parameters $g_k$ and $\Omega_{r,k}$ respectively characterize the coupling strength between the transmon and the $k$-th mode, and the drive strength of the probe field applied to $k$-th mode.
$\omega_{r,k}$ are generally far off-resonant from both $\omega_d$ and $\omega_d$, and thus, $g_k$ and $\Omega_{r,k}$ are effectively zeros in those modes.

The exceptional cases are the fundamental modes of unused CPWRs in Q1–Q3. For these modes, $g_k / |\omega_{r,k} - \omega_q| \approx 0.1 $ are satisfied, and thus, they noticeably renormalize the parameters of the transmons.
Hence, we take them into consideration when extracting the system parameters and calibrating $P_r$ and $\widetilde{V}_{\textup{JJ}}$ for more accuracy.
The validity of the semiclassical approximation also takes into account the effects of the resonator modes lying close to the transmon frequencies for Q1–Q3.
These effects are discussed in the following subsection with a subtitle `The semiclassical approximation'.

\subsectionntoc{Reduced Hamiltonian model}
For a few lowest energy levels, we can reduce the system Hamiltonian to
\begin{equation}
 \begin{aligned}
    \hat{H}_{\textup{low}} \approx ~ \widetilde{\omega}_{q} \hat{a}^\dagger \hat{a}+ \widetilde{\omega}_{r} \hat{b}^\dagger \hat{b}
    -\frac{A_{q}}{2} {\hat{a}^\dagger\hat{a}^\dagger\hat{a}\hat{a}}
     - \frac{A_{r}}{2} {\hat{b}^\dagger\hat{b}^\dagger\hat{b}\hat{b}}
     - 2A_{qr}\hat{a}^\dagger\hat{a}\hat{b}^\dagger\hat{b},
\label{eq_calibration}
\end{aligned}
\end{equation}
when neglecting the unused modes ($\hat{a}_k$) and the probe term. 
Here, we introduce a transmon ladder operator $\hat{b} = \frac{1}{\sqrt{2}}(\frac{E_J}{8E_C})^{\frac{1}{4}}\hat{\varphi} + i\frac{1}{\sqrt{2}}(\frac{8E_C}{E_J})^{\frac{1}{4}}\hat{n}$ to simplify the transmon part of the Hamiltonian.
Here, $\hat{a}$, $\hat{n}$, and $\hat{\varphi}$ are resonator field, transmon Cooper-pair number, superconducting phase operators, respectively.
$\widetilde{\omega}_{q,r}$ refer to the dressed transmon and resonator frequencies. $A_{q,r}$ correspond to the self nonlinearity of the transmon and resonator. $A_{qr}$ indicates the cross nonlinearity between them and $A_{qr}\approx \sqrt{A_q A_r}$ is satisfied.
We resolve $\widetilde{\omega}_{q,r}$ and $A_{q, qr}$ from the transmon two-tone spectroscopy. Furthermore, the resonator bare frequency $\omega_r$ is given from the resonator transmission in the strong probe limit.  
Using these quantities, we extract the system parameters appearing in Eq. (1) of the main text, following the calibration procedure described in Ref.~\cite{calibration}. See \textbf{Supplementary Note 7} for details on the parameter extraction procedure.
For Q1–Q3, the fundamental modes of the unused CPWRs are also included in $\hat{H}_{\textup{low}}$ when extracting the parameters (see \textbf{Supplementary Note 6 and 7} for circuit details).

\clearpage

\subsectionntoc{The semiclassical approximation}
This section provides further details on how our system can be simplified from Fig.1(a) to Fig.1(b) of the main text.
Recall $\hat{a}=\left< \hat{a} \right> + {\delta \hat{a}}$ from the main text. The mean field part $\left< \hat{a} \right>$ evolves as $\left< \hat{a} \right> = -i\frac{\Omega_r}{\kappa}e^{-i\omega_d t}$ when the resonator dissipation rate is $\kappa$ for the given $\Omega_r$.
In Eq. (1) of the main text, eliminating the probe term and plugging $\left< \hat{a} \right> = -i\frac{\Omega_r}{\kappa}e^{-i\omega_d t}$ into the transmon–resonator interaction term yields Eq. (2) of the main text with $\Omega_q=2g{\Omega_r}/{\kappa}$. The resonator coherent photon number is $N_r=|\left\langle \hat{a} \right\rangle|^2=(\frac{\Omega_r}{\kappa})^2$.  This ends up with the relation in Fig. 1(b) of the main text.

As noted in the previous subsection, the fundamental modes of the unused coplanar waveguide resonators (CPWRs) lie close to the transmon frequencies for Q1–Q3, with non-negligible coupling strengths.
To verify the validity of the semiclassical approximation in the presence of these modes, we measure the noise powers emitted from them for sufficiently large $N_r$ with which the bare resonator responses appear.
From the measured power spectra, the estimated noise photon numbers $\langle \delta \hat{a}_k^\dagger \delta \hat{a}_k \rangle$ of these modes, where ${\delta \hat{a}_k} = \hat{a}_k - \langle \hat{a}_k \rangle$, are found an order of magnitude smaller than $\langle \delta \hat{a}^\dagger \delta \hat{a} \rangle$.
We therefore conclude that the influence of ${\delta \hat{a}_k}$ on both the transmon dynamics and the total energy dissipation rate is negligible.

\clearpage

\section*{Supplementary Note 2 : Floquet formalism}
The Floquet formalism provides a framework for describing dynamics of quantum systems driven periodically in time. 
For a Hamiltonian \( \hat{H}(t) = \hat{H}(t+T) \) with period \( T = 2\pi/\omega_d \), the solutions of the Schrödinger equation $\hat{H}(t)\ket{\psi(t)}=i\hbar\partial_t\ket{\psi(t)}$ can be expressed as
\begin{equation}
\begin{split}
|\psi(t)\rangle = \sum_{i}c_i |\psi_i(t)\rangle,
\end{split}
\end{equation}
where
\begin{equation}
\begin{split}
|\psi_i(t)\rangle = e^{-i\epsilon_{i}^{\alpha} t/\hbar} |\phi_i^{\alpha}(t)\rangle
\end{split}
\end{equation}
are the Floquet states and $c_i$ are complex numbers that satisfy $\sum_{i}|c_i|^2 = 1$.
Here, \( |\phi_i^{\alpha}(t)\rangle \) are time-periodic Floquet modes that meet \( |\phi_i^{\alpha}(t)\rangle \) = \( |\phi_i^{\alpha}(t+T)\rangle \) and \( \epsilon_i^{\alpha} \) are corresponding quasi-eigenenergies. $\alpha$ refers to the mode numbers. 
This approach allows mapping the time-dependent problem to an effective time-independent eigenvalue equation,
\begin{equation}
\begin{split}
[\hat{H}(t) - i\hbar \partial_t] |\phi_{i}^{\alpha}(t)\rangle = \epsilon_{i}^{\alpha} |\phi_{i}^{\alpha}(t)\rangle,
\end{split}
\end{equation}
facilitating analysis of dynamics under time periodic driving.

\subsectionntoc{Floquet modes}

Floquet modes are the bases of solutions to the Schrödinger equation for a system with a time-periodic Hamiltonian. In the Floquet formalism, these solutions are expressed as a product of a periodic function and an exponential time-dependent phase factor, where the quasi-eigenenergies is a constant but only defined up to multiples of $\omega_d$.
Countable infinite Floquet modes exist for each energy levels of the Hamiltonian.
The Floquet mode can be decomposed as $\ket{\phi_i^{\alpha}(t)}= \sum_{k}\ket{i}^{k} e^{-ik\omega_d t}$,
and here $\ket{i}^{k}$ is a Fourier component of $\ket{\phi_i^{\alpha}(t)}$ at frequency $k\omega_d$, which also can be decomposed into eigenbases of the undriven system $\ket{j}$ like $\ket{i}^{k} = \sum_{j}c_{ij}^{k}\ket{j}$. Here, $\sum_{j}|c_{ij}^{k}|^2 = 1$ is satisfied for each $i$ and $k$.

In the main text, we specifically use $\ket{\phi_i^{\alpha}(t)} = e^{i\alpha\omega_d t}\ket{\phi_i^{0}(t)}$ and $\epsilon_{i}^{\alpha}=\epsilon_{i}^0+\alpha\hbar\omega_d$ ($\alpha \in \mathbb{Z}$) as Floquet modes and corresponding quasi-eigenenergies of $\hat{H}_q(t)$ with an Floquet order of $\alpha$.
Here, $\ket{\phi_i^{0}(t)}$ and $\epsilon_{i}^0$ are Floquet modes and corresponding quasi-eigenenergies in the first Brillouin zone.
We define $\ket{\phi_i(t)}=\ket{\phi_i^{0}(t)}$ and $\epsilon_{i}=\epsilon_{i}^0$ in the main text.
For a Floquet mode $\ket{\phi_i^{\alpha}(t)}$, we define $\overline{H}_q$ 
\begin{equation}
\begin{split}
    \overline{H}_q = \frac{\omega_d}{2\pi} \int_{0}^{2\pi/\omega_d} dt \bra{\phi_i^{\alpha}(t)}\hat{H}_q(t)\ket{\phi_i^{\alpha}(t)}.
\end{split}
\end{equation}
This is independent ${\alpha}$, and thus, we drop $\alpha$ in  \textbf{Methods}.

\subsectionntoc{Floquet-Markov master equation}
The Floquet–Markov master equation is a theoretical framework for describing the time evolution of an open quantum system driven by a time-periodic Hamiltonian and coupled to Markovian environments. It combines Floquet theory, which addresses systems with periodic Hamiltonians, and the Born–Markov approximation, which assumes weak system–Markovic bath coupling. This formalism enables the treatment of dissipative dynamics of open quantum systems under time-periodic driving.

As in $\textbf{Methods}$, we express a specific transmon dissipation mechanism as `$\Psi$' when the relevant dimensionless operators describing the transmon–bath interaction are $\hat{\Psi}^{(l)}$.
We label each operators with $l$. 
We introduce $J^{(l)}(\omega)$, the normalized bath spectra (NBS) of the corresponding operators.
Then, Floquet-Markov formalism yields $\Gamma_{ij,k}^{(\Psi)} =\sum_{l}{J^{(l)}(\Delta_{ij,k})|\hat{\Psi}^{(l)}_{ij,k}|^2}$,
where $\hat{\Psi}_{ij,k}^{(l)} = \frac{\omega_d}{2\pi} \int_{0}^{2\pi/\omega_d} dt  \bra{\phi_i(t)}\hat{\Psi}^{(l)}\ket{\phi_j(t)}e^{-ik\omega_d t}$ in the zero-temperature limit.
Cooperating all the mechanisms into consideration, $\Gamma_{ij,k}$ and $\Gamma_{ij}$ are given by
$\sum_{\textup{$\Psi$}}\Gamma_{ij,k}^{\textup{($\Psi$)}}$ and 
$\sum_{\textup{$\Psi$}}\Gamma_{ij}^{\textup{($\Psi$)}}$, respectively. Thereby, we can set a rate equation among the Floquet modes as below
\begin{equation}
\begin{split}
\label{eq_mm4}
   \frac{dp_j}{dt} = \sum_{i}p_{i}\Gamma_{ji} - \sum_{i}p_{i}\Gamma_{ij}.
\end{split}
\end{equation}
In the steady state, $dp_j/dt$ becomes zero, and therefore, $\Sigma_{i}p_{i}\Gamma_{ij} = \Sigma_{i}p_{i}\Gamma_{ji}$ is satisfied as described in the main text.

\clearpage




\section*{Supplementary Note 3 : Transmon–bath interactions}

\subsectionntoc{Coherent and incoherent quantum state transitions}
Let us model the system–bath coupling Hamiltonian responsible for inducing state transitions as below
\begin{equation}
\label{SB}
\hat{H}_{SB} = \lambda(t)\hat{\Psi}.
\end{equation}
Here, \( \lambda(t) \) denotes a time-dependent quantity with units of energy, and \( \hat{\Psi} \) is a dimensionless operator associated with a given transition mechanism \( \Psi \).
For certain mechanisms, such as quasiparticle generation (QPG), multiple operators \( \hat{\Psi} \) may be required, however, for simplicity, we restrict our discussion to a single operator. For monochromatic $\lambda(t)$, ($\lambda(t)=\lambda_0\cos(\omega_d t)$), $\hat{H}_{SB}$ induces coherent and reversible transitions.
In the perturbative regime, the transition probability between $i$ and $j$ states is given by 
\begin{equation}
P_{ij}=\frac{\lambda_0^2}{\hbar^2}\frac{\sin^2[(\omega_{ij}-\omega_d)]}{(\omega_{ij}-\omega_d)^2}|\bra{i}\hat{\Psi}\ket{j}|^2.
\end{equation}
Here, $\hbar\omega_{ij}$ is the energy difference of the $i$ and $j$ states of the system.

However, baths always induce incoherent noises on the systems, and thus, $\lambda(t)$ describing the system–bath interactions should be stochastic functions in time.
If the properties of the baths is not varying in time, then $\lambda(t)$ is stationary, which means that its power spectrum $S_{\lambda\lambda}(\omega)$ can be characterized as below
\begin{equation}
J_{\lambda\lambda}(\omega)
= \int_{-\infty}^{\infty} 
\langle \lambda(\tau)\lambda(0) \rangle \, e^{i\omega \tau}\, d\tau ,
\end{equation}
where the bracket $\langle \cdots \rangle$ denotes the ensemble average, which coincides with the time average for ergodic $\lambda(t)$.  
The system–bath interaction $\hat{H}_{SB}$ then induces incoherent transitions between states $i$ and $j$, with a transition probability $P_{ij}$ given by
\begin{equation}
P_{ij} = \frac{|\bra{i}\hat{\Psi}\ket{j}|^2}{\hbar^{2}}
\int_{0}^{t} J_{\lambda\lambda}(\omega) 
\frac{\sin^{2}\!\big[(\omega_{0}-\omega)t/2\big]}{(\omega_{0}-\omega)^{2}} 
\, d\omega
\simeq \frac{|\bra{i}\hat{\Psi}\ket{j}|^2}{\hbar^{2}}\, J_{\lambda\lambda}(\omega_{0})\, t .
\end{equation}
The above relation holds for sufficiently weak system–bath couplings.
The transition rate $\Gamma_{ij}$ is given by the time derivation of $P_{ij}$ as below.
\begin{equation}
\Gamma_{ij} = \frac{dP_{ij}}{dt}.
\end{equation}

\subsectionntoc{The quantum fluctuation-dissipation theorem on circuits}

\begin{figure*}
    \centering
    \includegraphics[width=0.8\linewidth]{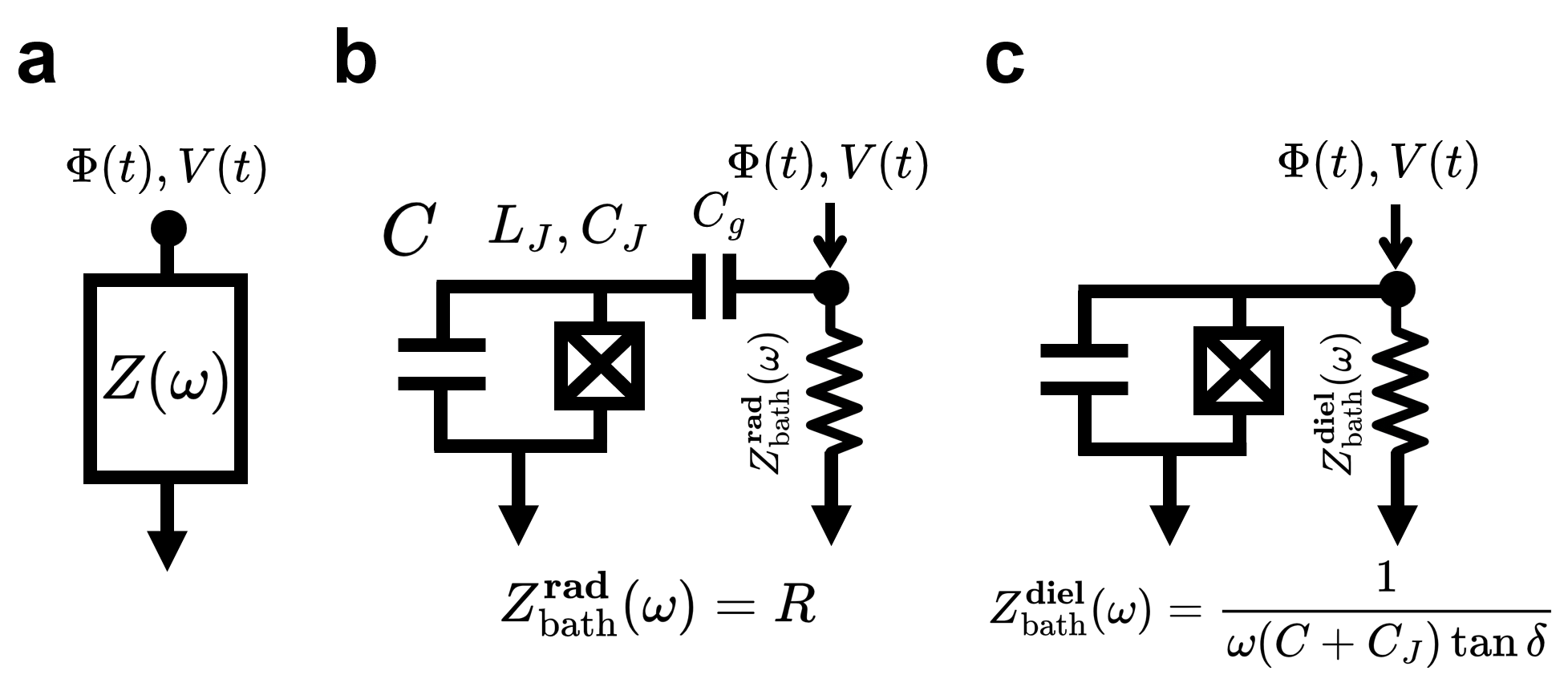}
    \caption{\textbf{Circuit diagram for illustrating the quantum fluctuation-dissipation theorem on transmons.} \textbf{(a)} A generalized circuit element of impedance $Z(\omega)$. In the diagram, $\Phi(t)$ and $V(t)$ indicate the quantum flux and voltage fluctuations as functions of time $t$. \textbf{(b)} A transmon capacitively coupled to Ohmic radiative bath ($Z^{\textbf{rad}}_{\textup{bath}}(\omega)=R$) represented by a resistor. $C_g$ is the coupling capacitance. $L_J$ and $C_J$ are inductance and capacitance of the Josephson junction of the transmon. \textbf{(c)} A transmon coupled to dielectric bath ($Z^{\textbf{diel}}_{\textup{bath}}(\omega)$). $C$ is the capacitance of the transmon pad. $\tan\delta$ indicates the dielectric loss tangent.}
    \label{fig:dissipation_1}
\end{figure*}

Fig.~\ref{fig:dissipation_1}(a) shows an arbitrary circuit element with an impedance $Z(\omega)$.
The quantum fluctuation-dissipation theorem on circuits \cite{fdt} expects that the flux $\Phi(t)$ across the element follows
\begin{equation}
\langle \Phi(t)\,\Phi(0)\rangle
= \frac{\hbar}{2\pi} 
\int_{-\infty}^{+\infty} \frac{d\omega}{\omega}
\left[ \coth\!\left( \frac{\beta \hbar \omega}{2} \right) + 1 \right]
\mathrm{Re}\!\left( Z[\omega] \right) e^{-i\omega t}.
\end{equation}
Here, $\beta=1/k_{B}\Theta$. $k_{B}$ and $\Theta$ respectively indicate Boltzmann constant and the temperature.
The spectrum of $\Phi$, $J_{\Phi\Phi}(\omega)$, is given by
\begin{equation}
J_{\Phi\Phi}(\omega)
= \int_{-\infty}^{+\infty} dt \,
\langle \Phi(t)\,\Phi(0) \rangle \, e^{i\omega t}=\frac{\hbar}{\omega}
\left[ \coth\!\left( \frac{\beta \hbar \omega}{2} \right) + 1 \right]
\mathrm{Re}\!\left( Z[\omega] \right).
\end{equation}
$V(t)$ represents the voltage across the element. $V(t)=\dot{\Phi}(t)$ is satisfied by definition. 
Then, the spectrum of the voltage, $S_{VV}(\omega)$, reads
\begin{equation}
J_{VV}(\omega)
= \int_{-\infty}^{+\infty} dt \,
\langle \dot{\Phi}(t)\,\dot{\Phi}(0) \rangle \, e^{i\omega t} =
{\hbar\omega}
\left[ \coth\!\left( \frac{\beta \hbar \omega}{2} \right) + 1 \right]
\mathrm{Re}\!\left( Z[\omega] \right).
\end{equation}
In the zero temperature limit($\beta\hbar\omega\gg1$), we obtain the below relations.
\begin{equation}
\begin{split}
\label{eq_fdt}
J_{\Phi\Phi}(\omega) &= \frac{2\hbar}{\omega}\mathrm{Re}\!\left( Z[\omega] \right), \\
J_{VV}(\omega) &= {2\hbar\omega}\mathrm{Re}\!\left( Z[\omega] \right).
\end{split}
\end{equation}
Based on these relation, we can readily derive
\begin{equation}
\begin{split}
\label{eq_fdt2}
J_{II}(\omega) = {2\hbar\omega} \mathrm{Re}\!\left( Y[\omega] \right),
\end{split}
\end{equation}
where $J_{II}(\omega)$ refers to the spectrum of $I(t)$, the current across the element.
$Y[\omega]$ in Eq.~\ref{eq_fdt2} is the admittance of the element.
These relations are essential for describing the dissipative dynamics of transmons coupled to noise baths.

Fig.~\ref{fig:dissipation_1}(b–c) depicts the coupling mechanisms between a transmon and either the Ohmic radiative bath (b) or the dielectric bath (c). In these configurations, the transmon is capacitively coupled to the radiative bath (b) and galvanically to the dielectric bath (c). We can derive $\hat{H}_{SB}$ between the transmon and baths from these pictures in the weak system–bath coupling regime. The corresponding $\lambda(t)$ should be simple functions of $V(t)$, $\Phi(t)$, or $I(t)$. The relations in Eq.~\ref{eq_fdt} have important roles for offering their power spectra.

\clearpage

\subsectionntoc{Noise bath spectra for transmons}

\subsubsectionnotoc{Radiation}
The Hamiltonian that describes the system and radiative bath reads
\begin{equation}
\begin{split}
    \hat{H}_{SB}^{(\textbf{rad})} &= -\frac{4 E_C C_g V(t)}{e}\hat{n}.
\end{split}
\end{equation}
Here, the spectrum of $V(t)$ follows Eq.~\ref{eq_fdt}. 
In this case, what correspond to $\lambda(t)$ and $\hat{\Psi}$ in Eq.~\ref{SB} are $-\frac{4E_C C_g V(t)}{e}$ and $\hat{n}$, respectively.
When defining $Q_{\textbf{rad}}^{-1}$ by $4\pi(\frac{R}{R_K})(\frac{C_g}{C+C_J})^2$, this leads to
\begin{equation}
\begin{split}
    &J^{\textbf{(rad)}}(\omega) =  \frac{\omega}{Q_{\textbf{rad}}}.  
\end{split}
\end{equation}
Here, we introduce Klitzing constant $R_K=g_K^{-1}=h/e^2$ in $Q_{\textbf{rad}}$.

\begin{figure*}
    \centering
    \includegraphics[width=0.8\linewidth]{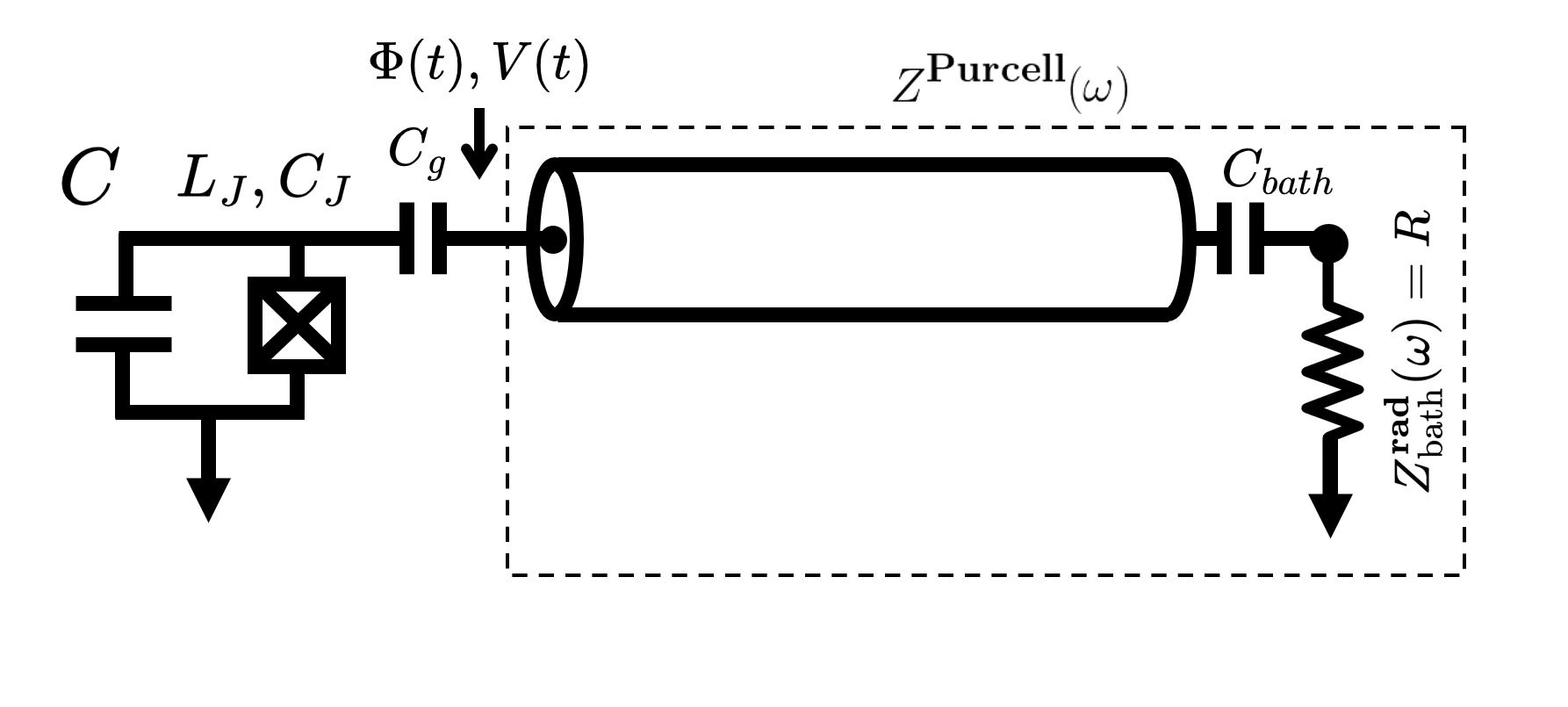}
    \caption{\textbf{Circuit model to describe the Purcell effect.} A finite length of a trasmission line is capacitively coupled to a transmon with a capacitance $C_g$. The trasmission line is also capacitively coupled to  Ohmic radiative bath ($Z^{\textbf{rad}}_{\textup{bath}}(\omega)=R$) with a capacitance $C_{bath}$. The resistor represents the Ohmic radiative bath of a resistance $R$. Here, $L_J$ and $C_J$ are inductance and capacitance of the Josephson junction of the transmon. $C$ is the capacitance of the transmon pad. $Z^{\textbf{Purcell}}_{\textup{bath}}(\omega)$ represents the impedance of the dashed box. 
    $\Phi(t)$ and $V(t)$ indicate the quantum flux and voltage fluctuations where the arrow points out.}
    \label{fig:dissipation_2}
\end{figure*}

\begin{figure*}
    \centering
    \includegraphics[width=0.8\linewidth]{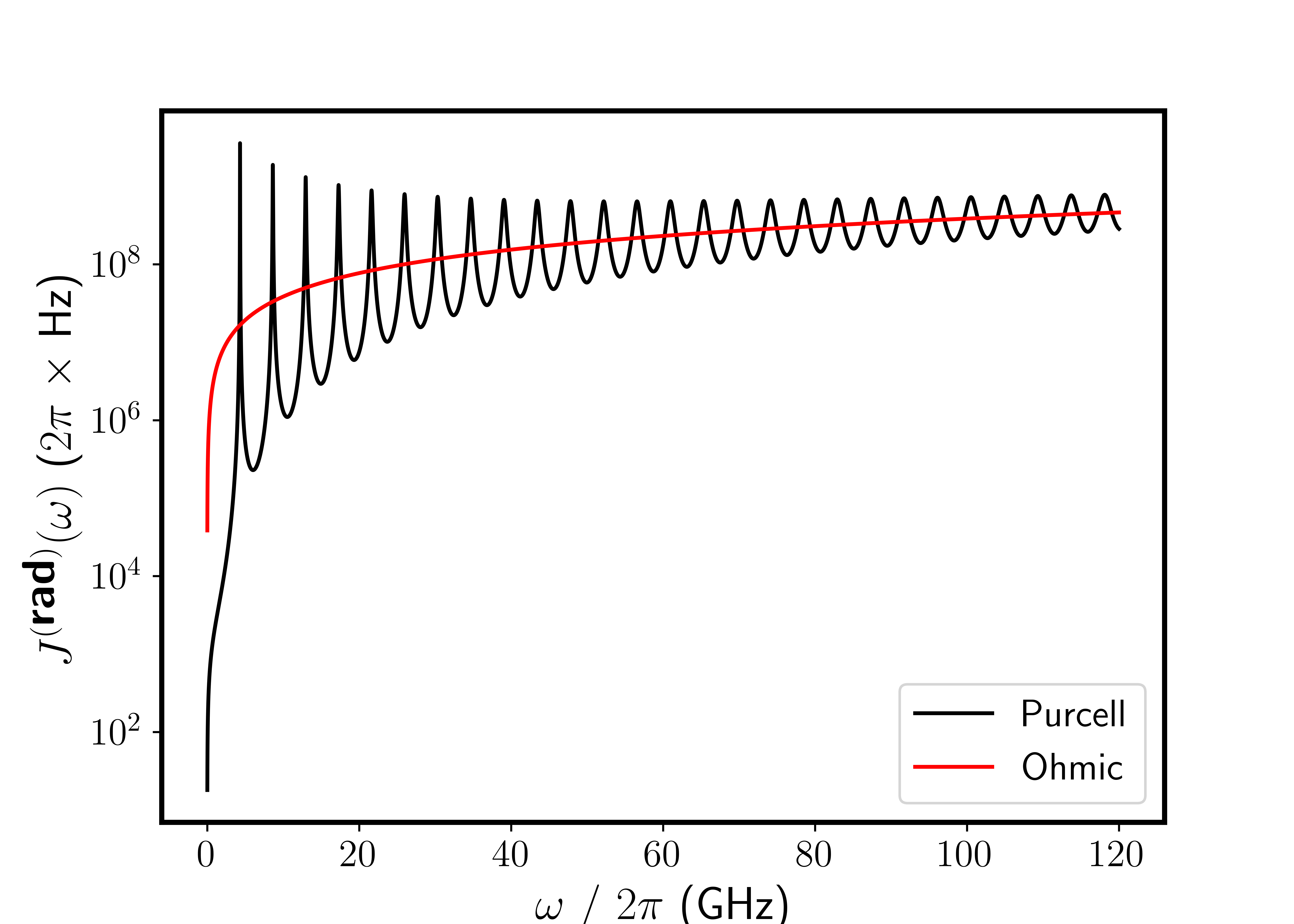}
    \caption{\textbf{Ohmic and Purcell models for $J^{\textbf{(rad)}}(\omega)$.} The Purcell model is obtain from $Z^{\textbf{Purcell}}(\omega)$ in Fig.~\ref{fig:dissipation_2}, with $C_{g}=40$ fF, $C_{bath}=50$ fF, and $R=50~\Omega$. For the Ohmic model, we properly set $R$ such that $J^{\textbf{(rad)}}(\omega)$ behaves similarly with that of the Purcell model in the large $\omega$ limit.}
    \label{fig:purcell}
\end{figure*}

The readout resonator and other omitted resonator modes coupled to the transmon can modify the radiative environment via the Purcell effect.
Fig.~\ref{fig:dissipation_2} illustrates a representative example: a transmon capacitively coupled to a transmission-line resonator, which in turn is capacitively coupled to a radiative bath modeled by a resistor.
In this configuration, the flux and voltage, $\Phi(t)$ and $V(t)$, at the point indicated by the arrow are governed by the effective impedance $Z^{\textup{Purcell}}(\omega)$ rather than by the simple Ohmic model.
Figure \ref{fig:purcell} compares the resulting $J^{\textbf{(rad)}}(\omega)$ for the Ohmic and Purcell models. 
%
%
Although the two models differ substantially for frequencies $\omega$ near the resonant modes of the transmission-line resonator, their overall behavior becomes similar in the large-$\omega$ limit. The integrals of $J^{\textbf{(rad)}}(\omega)$ in both models agree to within an order of magnitude.
Therefore, the Purcell effect introduces neither an order-of-magnitude change in $\mathcal{T}$ nor any noticeable change in $p_i$.
\\*
\subsubsectionnotoc{Dielectrics}

The Hamiltonian that describes the system and dielectric bath reads
\begin{equation}
\begin{split}
    \hat{H}_{SB}^{(\textbf{diel})} &= - I(t) \hat{\phi} = -  \frac{\hbar I(t)}{2e} \hat{\varphi}.\end{split}
\end{equation}
Here, the spectrum of $I(t)$ follows Eq.~\ref{eq_fdt2}. Here, what correspond to $\lambda(t)$ and $\hat{\Psi}$ in Eq.~\ref{SB} are $-\frac{\hbar I(t)}{2e}$ and $\hat{\varphi}$, respectively. 
Then, this readily yields
\begin{equation}
\begin{split}
    &J^{\textbf{(diel)}}(\omega)= \frac{\hbar\omega^2}{4E_C Q_{\textbf{diel}}(\omega)}.
\end{split}
\end{equation}
Here, we define $Q_{\textbf{diel}}(\omega)$ as the inverse of the loss tangent $\tan\delta$.
Please also recall that we hypothetically set $1/Q_{\textbf{diel}}(\omega)= (1/Q_{\textbf{diel}})e^{-\omega/\omega_{\textbf{diel}}^c}$ in the main text. $\omega_{\textbf{diel}}^c$ is the high-frequency cutoff of the dielectric bath spectrum.
\\*
\subsubsectionnotoc{Quasiparticle generation}
The transitions among Floquet modes with $\hbar\Delta_{ij,k} > 2\Delta_{\textup{Al}}$ can occur with the quasiparticle generation (QPG) across JJs. 
The description of bath model regarding QPG lacks the clear circuit representation as in radiation and dielectrics cases.
Ref.~\cite{Houzet-PRL-2019,QPG-1,QPG-2} illustrate the QPG processes quantitatively. We adopt the model therein in this work. 
The system–bath interaction in this case reads
\begin{equation}
\hat{H}_{SB}^{(\textbf{QPG})} = \tau \sum_{l,r,s} 
\Bigg[
\left( u_r v_l + v_r u_l \right) \cos\left( \frac{\hat{\varphi}}{2} \right)
+ i \left( u_r v_l - v_r u_l \right) \sin\left( \frac{\hat{\varphi}}{2} \right) 
\Bigg] \hat{\gamma}_{rs}^{\dagger} \hat{\gamma}_{ls}^{\dagger} + \text{h.c.}
\end{equation}
Here, $l$, $r$, and $s$ stand for the left and right sides of the JJ and the electron spin.
$\tau$, $u_{r,l}$ and $v_{r,l}$ are a tunneling amplitude and BCS coherence factors of each side of the JJ, respectively.
$\hat{\gamma}_{ls}$ and $\hat{\gamma}_{rs}$ mean the annihilation operators for quasiparticles on each side of the JJ.
The dimensionless operators involved in the system–bath interaction are $\sin{(\hat{\varphi}/2)}$ and $\cos{(\hat{\varphi}/2)}$.
The quantum fluctuations of $\hat{\gamma}_{ls}$ and $\hat{\gamma}_{rs}$ induce stochastic time-varying noises in $\hat{H}_{SB}$, corresponding $\lambda(t)$ in Eq.~\ref{SB}.
For the spectra of this noise, we adopt the model in Ref.~\cite{Houzet-PRL-2019,QPG-1,QPG-2}.
$J^{(\textbf{QPG}\pm)}(\omega)$ presented in \textbf{Methods} without cutoff factors shows the forms of the spectra.

Let us define $\sigma^{\pm}_{\textup{ideal}}(\omega)$ as ideal QPG conductance.
Neglecting the imaginary part of $\sigma_{\pm}^{\textup{ideal}}(\omega)$ and introducing an effective inductance $L_{\textup{eff}}^{\pm}$ in series with $\sigma_{\pm}^{\textup{ideal}}(\omega)$ yields the total conductance
\begin{equation}
\begin{split}
\sigma^{\pm}(\omega)
= \sigma^{\pm}_{\textup{ideal}}(\omega)
\frac{1}{1 - \frac{i\omega}{\omega^c_{\textbf{QPG}\pm}}}.
\end{split}
\end{equation}
Here, $\omega^c_{\textbf{QPG}\pm} = \big(\sigma^{\pm}_{\textup{ideal}}(\omega)L^{\pm}_{\textup{eff}}\big)^{-1}$ is satisfied. Utilizing the relation $\mathrm{Re}[\sigma_{\textup{QPG}}^{\pm}(\omega)] = \frac{\pi g_K}{\omega} J^{(\textbf{QPG}\pm)}(\omega)$, we obtain
\begin{equation}
\begin{split}
J^{(\textbf{QPG}\pm)}(\omega)
= \frac{J^{(\textbf{QPG}\pm)}_{\textup{ideal}}(\omega)}
{1 + \left(\frac{\omega}{\omega^c_{\textbf{QPG}\pm}}\right)^2 }.
\end{split}
\end{equation}
In \textbf{Methods}, we assume $\omega_{\textbf{QPG}+}^c = \omega_{\textbf{QPG}-}^c$.
As mentioned in \textbf{Methods}, $J^{(\textbf{QPG}-)}(\omega)$ contributes negligibly to dissipation compared with $J^{(\textbf{QPG}+)}(\omega)$, so we focus on the effects from $J^{(\textbf{QPG}+)}(\omega)$ throughout the main text and supplementary information. 
We define $\sigma(\omega) = \mathrm{Re}[\sigma^{+}(\omega)]$ and $\omega_{\textbf{QPG}}^{c} = \omega_{\textbf{QPG}+}^{c}$ in the main text.
\\*
\subsubsectionnotoc{Tunneling of non-equilibrium quasiparticles}
In practice, the JJs of superconducting qubits host non-equilibrium quasiparticles above the superconducting gap. Drives can also induce the tunneling events such non-equilibrium quasiparticles across the JJs, which can contribute to the loss at the driven transmons.
They are typically quantified as
\begin{equation}
x_{\textup{qp}} = \frac{n_{\textup{qp}}}{2 \nu \Delta_{\textup{Al}}},
\end{equation}
which is the quasiparticle density normalized by the Cooper-pair density.
Here, $n_{\textup{qp}}$ denotes the quasiparticle number, and $\nu$ is the
density of states per spin direction.
For typical superconducting–qubit devices, $x_{\textup{qp}}$ is generally
reported to lie in the range $10^{-8}$–$10^{-5}$.
Since we employ sufficient infrared shielding for our devices (see \textbf{Supplementary Note 6}), we believe that $x_{\textup{qp}}$ should not largely exceed the typical range. Meanwhile, it has been known that $x_{\textup{qp}}$ follows the thermal distribution above temperatures larger than 200 mK, satisfying
\begin{equation}
x_{\textup{qp}}=\sqrt{\frac{2\pi k_B \Theta}{\Delta_{\textup{Al}}}}e^{-\Delta_{\textup{Al}}/(k_B \Theta)}.
\end{equation}

$x_{\textup{qp}}$ is approximately $10^{-5}$ with $\Theta=200$ mK and $3\times10^{-5}$ with $\Theta=300$ mK. Assuming that the non-equilibrium quasiparticle density is $10^{-5}$ in our transmons, it becomes three times larger when $\Theta=300$ mK. 
If the effect of non-equilibrium quasiparticles were significant, we would also observe a corresponding change in the nonlinear dissipation of the resonators at the bare responses.
Nonetheless, we cannot identify any noticeable change in the nonlinear dissipation with increasing the device temperature up to 300 mK.
Eventually, this leads to the conclusion that drive-induced tunneling events of non-equilibrium quasiparticles contribute negligibly to the total dissipation at the driven transmons.

\clearpage

\section*{Supplementary Note 4 : Methods on Floquet numerical calculations}
Floquet-based numerical simulations in this work are performed using QuTiP (version~4). 
The Floquet modes $\ket{\phi_i^{0}(0)}$ and quasi-energies $\epsilon_i^0$ are obtained via the \textsf{floquet\_modes} function, and their time evolution over one drive period is evaluated using \textsf{floquet\_modes\_table} function. 
The matrix elements $\hat{\Psi}_{ij,k}^{(l)}$ and transition rates $\Gamma_{ij,k}^{\Psi}$ are then calculated with \textsf{floquet\_master\_equation\_matrix} and \textsf{floquet\_master\_equation\_rate}, respectively. 
After obtain the transition rates $\Gamma_{ij}=\sum_{k}\Gamma_{ij,k}$ are obtained, the steady-state populations $p_i$ are computed using the \textsf{floquet\_markov\_mesolver} function. 
To avoid excessive memory usage and enable systematic debugging, each step is executed independently.

\subsectionntoc{Parameters for the numerical calculations}

\begin{center}
\begin{table}
 \begin{tabular}{||c||c||c||} 
 \hline
 Symbols & Definitions & Values \\ [0.5ex] 
 \hline\hline
 $\mathcal{D}$         & Hilbert space dimension & 401$^\dagger$ \\ 
 \hline
 $k_{\textup{max}}$   & Maximum value of $k$ in $\hat{\Psi}_{ij,k}$ and $\Gamma_{ij,k}^{(\Psi)}$ & 200 \\ 
 \hline
 $N_T$          & Number of time steps in $\hat{\Psi}_{ij,k}$ and $\overline{H}_q$ & 20001 \\ 
 \hline
 $N_t$          & Number of time points in \textsf{floquet\_modes\_table} & 2001 \\ 
 \hline
 \end{tabular}
 \caption{\label{tab:numerical_params} 
 \textbf{Parameters used in the Floquet simulation.}  
 Key numerical parameters for the Floquet calculations that critically affect accuracy of the numerical results.  $^\dagger$~For the radiative case, we limit $\mathcal{D}$ by 201. See the text for the reason.}
\end{table}
\end{center}

Tab.~\ref{tab:numerical_params} summarizes the key numerical parameters used in the Floquet calculations.
The used Hilbert-space dimension is 401, denoted by $\mathcal{D}$
From Fig.~2(d) and Fig.~4(c) in the main text, $\mathcal{D}=401$ is clearly sufficient to ensure the accuracy of our calculations.
We restrict to $\mathcal{D}=201$ when computing the radiative loss.
Among all quantities, $\hat{\Psi}_{ij,k}$ and $\Gamma_{ij,k}^{(\Psi)}$ with $\Psi=\textbf{rad}$ are the most sensitive to numerical errors. Increasing $\mathcal{D}$ beyond 201 would require substantially larger values of $N_t$ and $N_T$ to maintain sufficient numerical accuracy.
This practical limitation explains why the numerical data for the radiative loss in Fig.~4 and Fig.~5 of the main text extends only up to 120 GHz.
$k_{\textup{max}}$ is the maximum value of $k$ in $\hat{\Psi}_{ij,k}$ and $\Gamma_{ij,k}^{(\Psi)}$.
We set this by 200 throughout the numerical calculations presented in this work.
$N_t$ means the number of time points in \textsf{floquet\_modes\_table}. The higher $N_t$ is, the more accurate the computed $|\phi_i^{\alpha}(t)\rangle$ becomes.
$N_T$ indicates the number of time steps when performing the time integrals in computing $\hat{\Psi}_{ij,k}$ and $\overline{H}_q$.

\subsectionntoc{Population $p_i$ and $\mathcal{T}$}
For a given set of $\Gamma_{ij}$, the steady-state population $p_i$ can be
computed using Eq.~\ref{eq_mm4}. This procedure is equivalent to solving the
rate equations, and the function \textsf{floquet\_markov\_mesolve} performs
this task. We evolve the system for a sufficiently long time such that further
extension of the evolution produces no noticeable change in $p_i$. 
Meanwhile, we find that $p_i$ for $i > N_{\textup{ch}}$ is mainly determined by numerical errors. 
Thus, we exclude the components with $p_i$ for $i > N_{\textup{ch}}$ when calculating $\mathcal{T}$.
Furthermore, we find that  $\sum_{j,k}\Gamma^{\textbf{(QPG)}}_{ij,k} \Delta_{ij,k}$ is insensitive to $i$. Readily, the calculated $\mathcal{T}$ becomes insensitive to $p_i$ for a given
$\Omega_q$. 
Therefore, any potential numerical errors in $p_i$ do not lead to
an order-of-magnitude change in $\mathcal{T}$.
\\*
\subsectionntoc{Validity of the numerical parameters}
We confirm that increasing $k_{\textup{max}}$, $N_t$, or $N_T$ beyond the chosen values does not produce any significant change in the calculated $\mathcal{T}$. This demonstrates that the parameters used in this work are sufficiently large to ensure numerical accuracy.
Supporting numerical results for this claim are provided in \textbf{Supplementary Note 5}.
\clearpage

\section*{Supplementary Note 5 : Results of Floquet numerical calculations}

\subsectionntoc{Transition matrix}

\begin{figure*}
    \centering
    \includegraphics[width=0.4\linewidth]{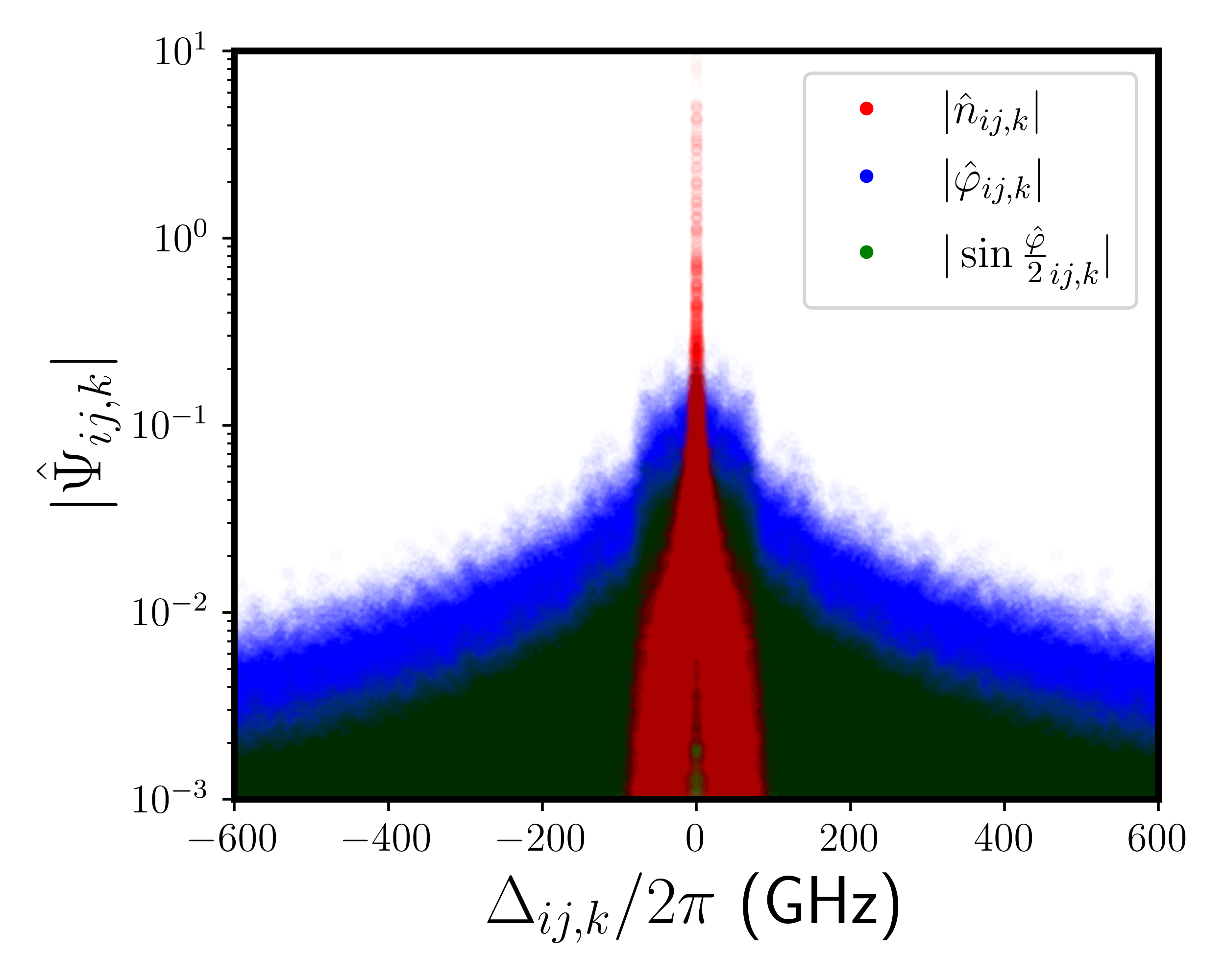}
    \caption{\textbf{Transition matrices.} Calculated $|\hat{\Psi}_{ij,k}|$ for each dissipation mechanisms with respect to $\Delta_{ij,k}$. In the calculation, $\Omega_q/2\pi$ is set by 40 GHz. Only the elements with $i,j\leq N_{\textup{ch}}\approx50$ are presented that dominantly affect the transmon dynamics.}
    \label{fig:transition_matrix}
\end{figure*}

Fig.~\ref{fig:transition_matrix} presents the calculated $|\hat{\Psi}_{ij,k}|$ as a function of $\Delta_{ij,k}$ for $\hat{n}$, $\hat{\varphi}$, and $\sin(\hat{\varphi}/2)$ at $\Omega_q/2\pi = 40$~GHz.
These operators are relevant to $\Gamma_{ij,k}^{(\textbf{rad})}$, $\Gamma_{ij,k}^{(\textbf{diel})}$, and $\Gamma_{ij,k}^{(\textbf{QPG})}$, respectively.
For $\hat{\varphi}$ and $\sin(\hat{\varphi}/2)$, which are associated with non-radiative transitions, the matrix elements $\hat{\Psi}_{ij,k}$ remain significant even for large $\Delta_{ij,k}$, reflecting higher-order photon processes.
In contrast, for $\hat{n}$, which is relevant to radiative transitions, $|\hat{n}_{ij,k}|$ decreases rapidly with increasing $|\Delta_{ij,k}|$.
For $\Delta_{ij,k}/2\pi > 100$~GHz, $|\hat{n}_{ij,k}|$ becomes negligible. 
In this regime, the computed values are dominated by numerical errors. Accordingly, we neglect $\hat{n}_{ij,k}$ for $|\Delta_{ij,k}| > 2\pi\times 100$~GHz when computing $\Gamma_{ij,k}^{(\textbf{rad})}$.

\subsectionntoc{Transition rates}

\begin{figure*}
    \centering
    \includegraphics[width=0.8\linewidth]{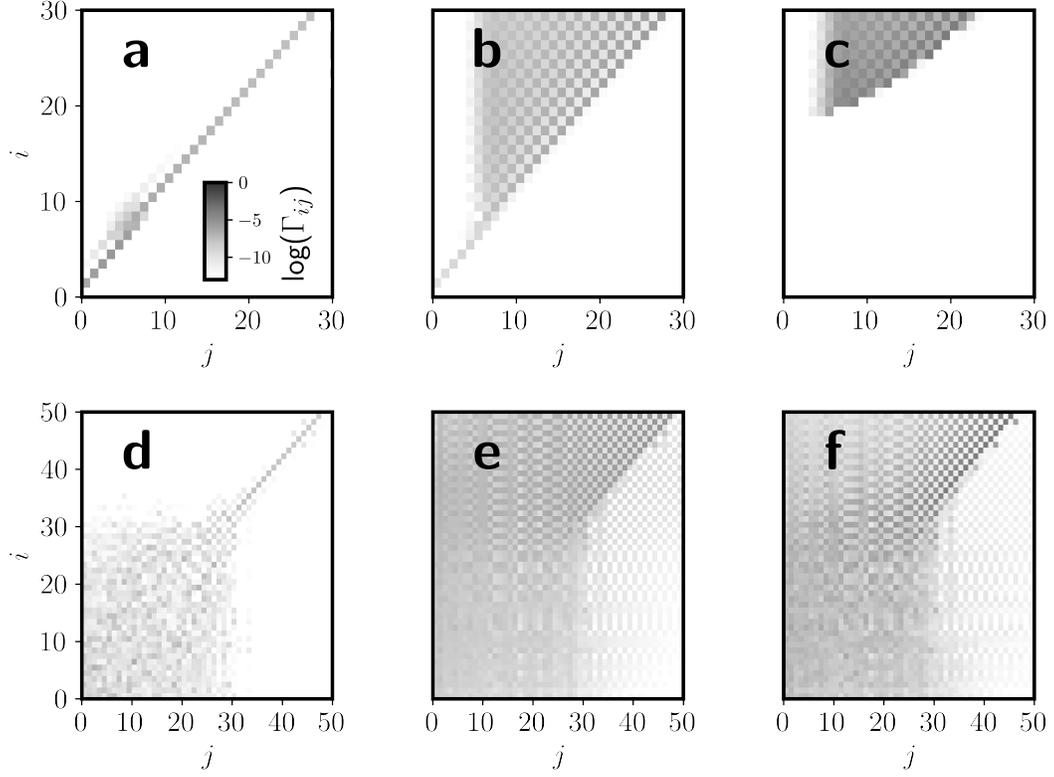}
    \caption{\textbf{Transition rates.} Calculated $\Gamma_{ij}$ for each dissipation mechanism. Each plot shows the results for the radiative, dielectric, and QPG mechanisms, from left to right. The bath parameters used in the calculation are $Q_{\textbf{rad}}=3830$, $Q_{\textbf{diel}}=4.8\times10^5$, $\omega_{\textbf{diel}}^c/2\pi=1$ THz, and $\omega_{\textbf{QPG}}^c/2\pi=17$ GHz. $Q_{\textbf{rad}}$ represents the approximate lower bound.
    \textbf{(a–c)} $\Omega_q=0$ in the calculations.
    \textbf{(d–f)} $\Omega_q/2\pi=20$ GHz in the calculations.}
    \label{fig:transition_rates}
\end{figure*}

Fig.~\ref{fig:transition_rates} presents the calculated $\Gamma_{ij}^{(\textbf{rad})}$, $\Gamma_{ij}^{(\textbf{diel})}$, and $\Gamma_{ij}^{(\textbf{QPG})}$, from left to right.
Fig.~\ref{fig:transition_rates}(a–c) and Fig.~\ref{fig:transition_rates}(d–f) show the results with $\Omega_q=0$, and $\Omega_q/2\pi=20$ GHz, respectively.
See the caption for the details on the bath parameters used in the calculations.

In Fig.~\ref{fig:transition_rates}(a–c), we confirm the radiative transition rates among the unconfined states ($i,j<7$) are dominant.
For the QPG mechanism, the transition rates among the unconfined states are even zero since the transition frequencies $\Delta_{ij,k}$ among these states are less than $2\Delta_{\textup{Al}}/h$.
For the transitions among the unconfined states ($i,j\geq7$), the situation has been entirely reversed. Dielectric and QPG mechanisms yield much larger $\Gamma_{ij}$ for $i,j\geq7$.
Furthermore, these mechanisms induce strong transitions among the distant energy levels ($|i-j|\gg 1$ and $\Delta_{ij,k}\gg \omega_d$).   

Under sufficiently large $\Omega_q$, the confined and unconfined states are hybridized into the chaotic Floquet modes.
Consequently, the effects of the large transition rates among the unconfined states dominate the transition rates among the chaotic Floquet modes.
This explains why the non-radiative transition rates, $\Gamma_{ij}^{(\textbf{diel})}$ and $\Gamma_{ij}^{(\textbf{QPG})}$, are far larger than $\Gamma_{ij}^{(\textbf{rad})}$ in Fig.~\ref{fig:transition_rates}(d–f). 
Our investigation presented in Fig.~\ref{fig:transition_rates} evidently shows that the substantial differences between the radiative and non-radiative $\Gamma_{ij}$ come from the dissipation of the unconfined states.

\clearpage

\subsectionntoc{Numerical accuracy in $\mathcal{T}$}

\begin{figure*}
    \centering
    \includegraphics[width=0.4\linewidth]{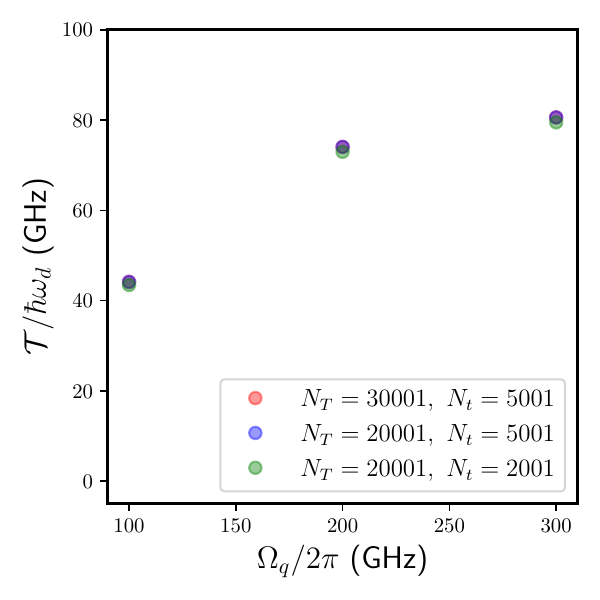}
    \caption{\textbf{Numerical simulation accuracy estimates.} We calculate $\mathcal{T}/\hbar\omega_d$ for $\Omega_q/2\pi = 100$, 200, and 300~GHz, and repeat the calculation for several different choices of key numerical parameters that affect the accuracy of the results. Only the QPG mechanism is considered in the calculations. $\omega_{\textbf{QPG}}^c/2\pi$ is set by 20 GHz. Other bath parameters are the same as those used in the main text.}
    \label{fig:accuracy}
\end{figure*}

In Fig.~\ref{fig:accuracy}, we present the calculated $\mathcal{T}/\hbar\omega_d$ for $\Omega_q/2\pi = 100$, 200, and 300~GHz using different numerical simulation parameters. We focus exclusively on the QPG case, which is the dominant contribution considered in this work.
We confirm that substantially increasing $N_T$ and $N_t$ does not lead to any significant increase in $\mathcal{T}$.
Therefore, the parameter values listed in Table~\ref{tab:numerical_params} provide sufficient numerical accuracy for evaluating $\mathcal{T}$.
In all calculations, $k_{\textup{max}}$ is fixed to 200, corresponding to a frequency range up to $200\times\omega_d$, which is sufficiently large to fully cover the QPG bath spectrum.

\subsectionntoc{$\mathcal{T}$ with various cutoff frequencies}
\begin{figure*}
    \centering
    \includegraphics[width=0.6\linewidth]{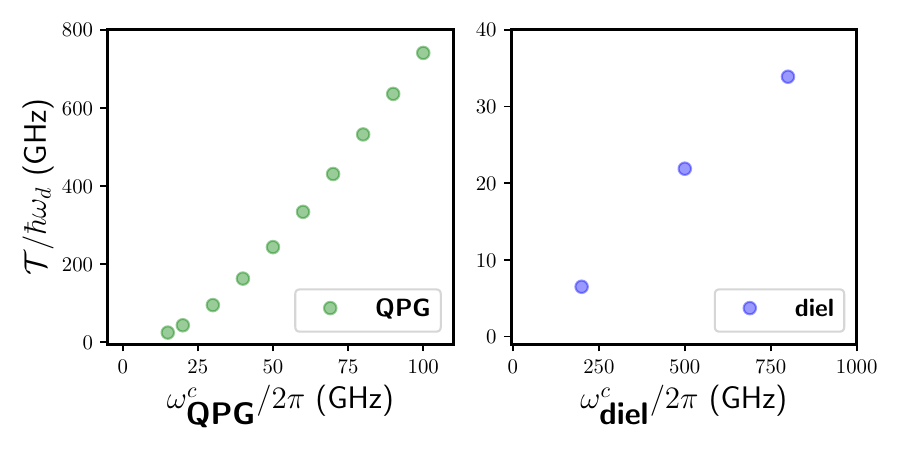}
    \caption{\textbf{Calculated $\mathcal{T}$ with various cutoff frequencies.} Left and right panels respectively indicate the calculated $\mathcal{T}/\hbar\omega_d$ based on QPG and dielectrics mechanisms while varying $\omega_{\textbf{QPG}}^c$ or $\omega_{\textbf{diel}}^c$. $\Omega_{q}/2\pi$ is set by 100 GHz. Other bath parameters are the same as those used in the main text.}
    \label{fig:nps}
\end{figure*}

In Fig.~\ref{fig:nps}, we present the calculated $\mathcal{T}/\hbar\omega_d$ with respect to $\omega_{\textbf{diel}}^c$ or $\omega_{\textbf{QPG}}^c$.
$\mathcal{T}$ in both cases monotonically increase with increasing $\omega_{\textbf{QPG}}^c$ or $\omega_{\textbf{diel}}^c$.
For QPG, $\mathcal{T}$ scales approximately quadratically with $\omega_{\textbf{QPG}}^{c}$ in the small-$\omega_{\textbf{QPG}}^{c}$ regime, while for larger $\omega_{\textbf{QPG}}^{c}$, $\mathcal{T}$ exhibits approximate linear scaling with $\omega_{\textbf{QPG}}^{c}$.
For dielectrics mechanism, $\mathcal{T}$ scales approximately linearly over the explored range.

\clearpage

\section*{Supplementary Note 6 : Experimental system}

\subsectionntoc{Devices under test}

\subsubsectionnotoc{Design}

\begin{figure*}
    \centering
    \includegraphics[width=0.9\linewidth]{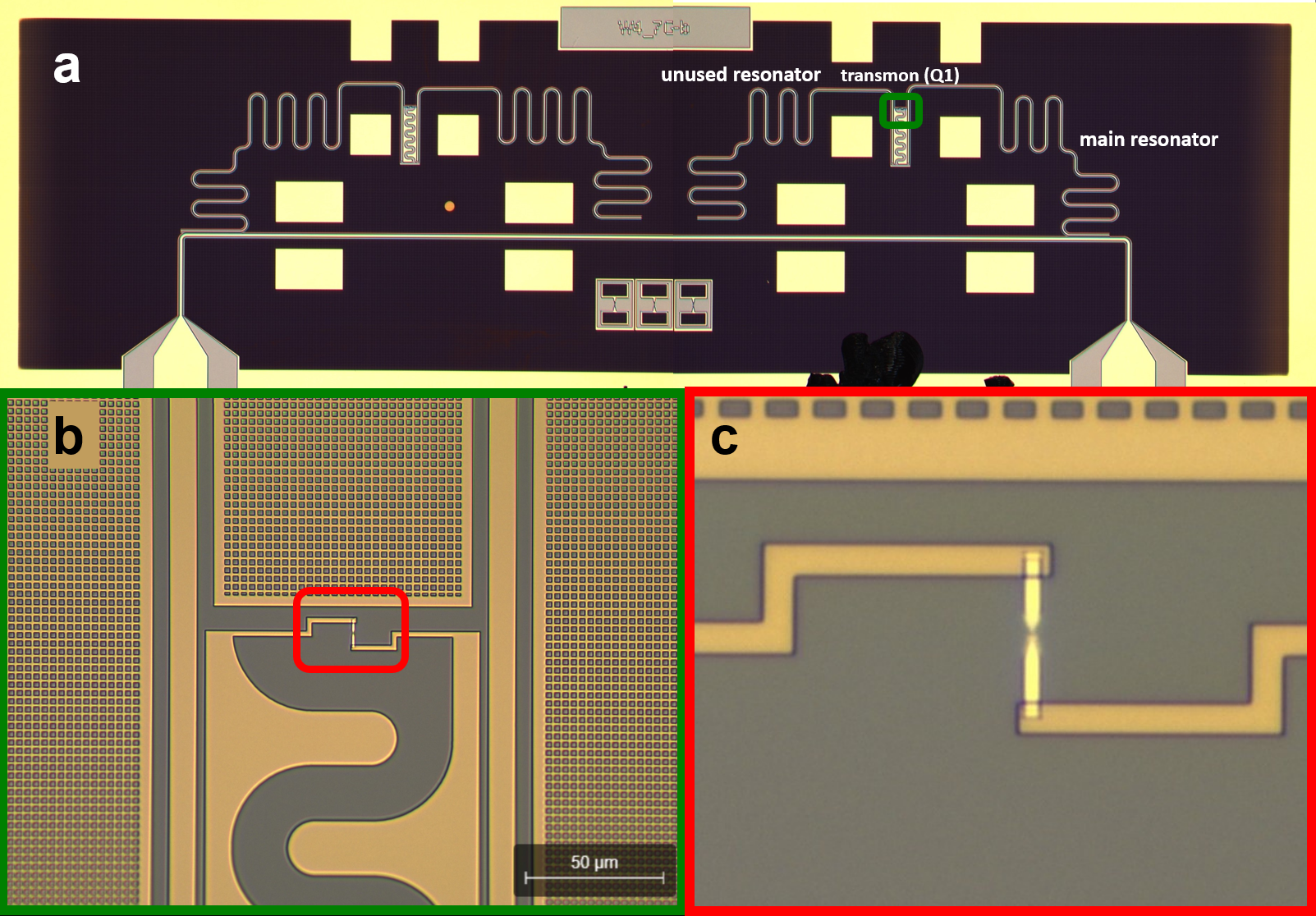}
    \caption{\textbf{Optical microscopy image of device under test.} \textbf{(a)} Overall image. \textbf{(b)} Magnified view of the transmon part. \textbf{(c)}  Magnified view of the Josephson junction part. Refer to the text for detailed descriptions.}
    \label{fig:device}
\end{figure*}

Fig.~\ref{fig:device} shows optical microscopy images of one of the devices under test.
Q1 is located in the right-hand-side of the presented device in Fig.~\ref{fig:device}(a). It is coupled to two coplanar waveguide resonators (CPWR), labeled main and auxiliary, respectively.
The main (unused) CPWR is strongly (weakly) coupled to the feedline.
In this work, we only use the fundamental mode of the main CPWR and neglect the higher modes of the main CPWR and all the modes of the unused CPWR.
For the device in the presented picture, the left-hand-side transmon is non-functional for reasons that remain unclear.
The rectangular features on the ground plane are designed for wirebonding. Wirebonds are subsequently made at these locations after the image is taken.
Q2 and Q3 are located in another device together. 
The designs for the transmons and appended circuits of Q2 and Q3 are identical with that of Q1. 
The Josephson junctions of Q1, Q2, and Q3 were fabricated simultaneously, and therefore, we expect their material properties are similar.
Q4 has a X-mon style circuit design and coupled to only one CPWR.

\subsubsectionnotoc{Fabrication}
The base layer was patterned on a 100 nm niobium titanium nitride (NbTiN) film on a 525 $\mu$m thick high-resistance silicon substrate. 
For the CPWRs, the center conductor width (W) and the gap between the center conductor and ground planes (S) are $10~\mu m$ and $6~\mu m$, which yield characteristic impedance of 60 $\Omega$.
For the feedlines, W and S are $20~\mu m$ and $6~\mu m$, which yield characteristic impedance of 50 $\Omega$.
The Josephson junctions of the transmons comprise of Al-AlOx-Al trilayer fabricated using double-angle shadow evaporation. The overlapped areas of the constituent Josephson junctions are approximately $100\times200~nm^2$, as confirmed in scanning electron microscopy.

\subsectionntoc{Cryostat}
\begin{figure*}
    \centering
    \includegraphics[width=0.9\linewidth]{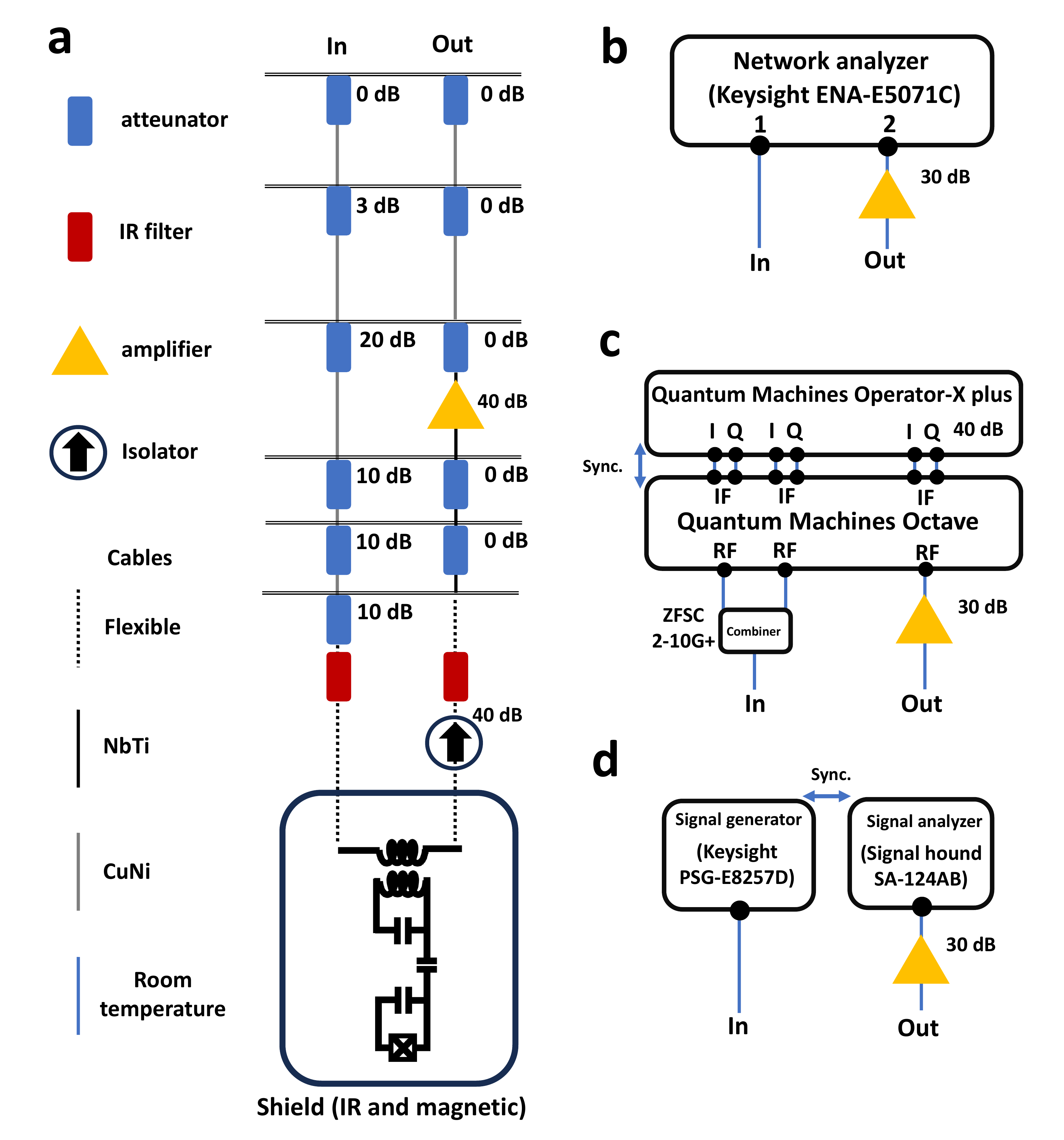}
    \caption{\textbf{Cryostat setup and measurement instruments.} \textbf{(a)} Schematic of the cryogenic wiring and electronics. \textbf{(b–d)} Room-temperature measurement setups for transmission spectroscopy, transmons two-tone spectroscopy, and resonator output power spectrum measurement, respectively.}
    \label{fig:cryo}
\end{figure*}

Fig.~\ref{fig:cryo}(a) shows the cryogenic setup used in the experiment.
The device is mounted on the mixing-chamber plate of a dilution refrigerator (Bluefors LD-400). 
Input and output signals are sent through separate lines, with 10/10/10/20 dB attenuators installed at the 4 K, 700 mK, 100 mK, and MXC stages of the input line, and a 40 dB isolator (LNF-ISISC) placed at the MXC stage of the output line.
A HEMT amplifier of approximately 40 dB gain (LNF-LNC) is positioned at the 4 K stage of the output line.
IR filters (Bluefors) are installed on each input and output lines.
The device is enclosed in radiation and magnetic shieldings made of copper and cryoperm, respectively.

\subsectionntoc{Measurement}

Fig.~\ref{fig:cryo}(b–d) shows the room-temperature measurement setups.
Transmission spectroscopy is performed using a two-port network analyzer (Keysight E5071C). 
The transmon characterization is performed by using an integrated quantum control platform (Quantum Machines OPX-plus and Octave).
For power spectrum measurements, we employ a signal generator (Keysight E8257D) and a spectrum analyzer (Signal Hound SA-124AB). 
The output signals from the refrigerator are amplified with a 30 dB power amplifier (Narda–MITEQ LNA-30-04000800) before being sent to the room-temperature instruments.

\clearpage

\section*{Supplementary Note 7 : Experimental details}

\subsectionntoc{Device parameters}

\begin{figure*}
    \centering
    \includegraphics[width=0.7\linewidth]{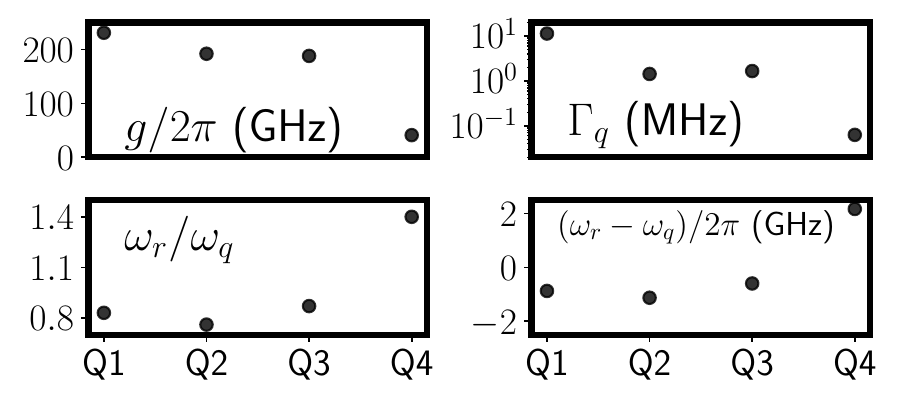}
    \caption{\textbf{Comparisons of some fixed parameters of the transmons (Q1–Q4) and corresponding readout resonators.}}
    \label{fig:params}
\end{figure*}

Table~\ref{tab:params} shows the key fixed parameters  of the transmons and readout resonators.
In Fig.~\ref{fig:params}, we compare come fixed parameters of Q1–Q4.
The $g$ of Q4 is one order-of-magnitude smaller than that of the others. This results in substantially smaller nonlinearity in $\kappa$, as remarked in Extended Data 1. More explanation is given in \textbf{Supplementary Note 9}.
We scan a wide range of $\Gamma_q$ but the data exhibit the similar upper bounds of $\omega^c_{\textbf{QPG}}$, as shown in Extended Data 1.

\begin{center}
\begin{table}
 \begin{tabular}{||c||c||c||c||c||c||} 
 \hline
 Symbols &  Definitions & Values (Q1) & Values (Q2) & Values (Q3)  & Values (Q4)\\ [0.5ex] 
 \hline\hline
 $\omega_r/2\pi$ & Resonator bare frequency (GHz) & 4.284  & 4.297 & 3.745 & 7.5474 \\ 
 \hline
 $\omega_q/2\pi$ & Transmon bare frequency$^\dagger$ (GHz) & 5.161  & 4.896 & 4.873 & 5.381\\ 
 \hline
 $E_J$ & Josephson energy ($h\cdot$GHz) & 14.24  & 13.01 & 13.02 & 12.06\\ 
 \hline
 $E_C$ & Charging energy ($h\cdot$MHz)&  259  & 257 & 254 & 310\\ 
  \hline
 $g$ & Transmon–resonator coupling ($h\cdot$MHz)& 231  &212 & 188 & 41 \\ 
 \hline
  $\Gamma_{q}$  & Transmon decay rate$^\dagger$ (MHz)   & 11.36  & 1.66 & 0.64 & 0.059 \\ 
 \hline
 $\kappa_{ex}$ & Resonator external decay (MHz) & 97.88  &  16.73&  27.38 & 6.43\\ 
 \hline
\end{tabular}
\caption{\label{tab:params} \textbf{System specification}. Key fixed parameters of the transmons (Q1–Q4) and corresponding readout resonators. Only the resonators used in the experiments are mentioned. $^\dagger$ : The transition frequency between the ground and the first excited states.}
\end{table}
\end{center}

\subsectionntoc{Device parameter extraction}
 
We compare $\hat{H}_{\textup{low}}$ in Eq.~\ref{eq_calibration} and $\hat{H}$ with $\Omega_r=0$ of the main text. We define $\textbf{E}_i(\hat{H})$ as $i$-th eigenenergy from the ground state of the Hamiltonian $\hat{H}$. The arguments of $\hat{H}_{\textup{low}}$ are experimentally obtained by the transmon two-tone spectroscopy. $\omega_r/2\pi=4.284$ GHz is given by the bare resonator response. We numerically optimize the system parameters $E_C$, $E_J$, $g$ such that the cost function $\sum_{i=0}^{2}(\textbf{E}_i(\hat{H})-\textbf{E}_i(\hat{H}_{\textup{low}}))$ is minimized. We take only the lowest three energy levels into consideration. We use `\textsf{scipy.minimize}' function for this procedure \cite{scipy}.

\clearpage

\subsectionntoc{Calibration}

\subsubsectionnotoc{Resonator probe powers $P_r$}
The resonator transmission exhibits $P_r$ dependence in the dressed frequency ($\widetilde{\omega}_r$) response as identified in Fig 1(c). We utilize this behavior to calibrate $P_r$. In this regime, the simplified Hamiltonian model $\hat{H}_{\textup{low}}$ in Eq.~\ref{eq_calibration} is available when $N_r$ is sufficiently low.
The model indicates the resonator frequency shift $\delta\widetilde{\omega}_r$ is approximated given by $-2A_rN_r$. We obtain $A_r$ from the transmon two-tone spectroscopy.
Together with the relation $N_r=\frac{2P_r}{\hbar\omega_d}\frac{\widetilde{\kappa}_{ex}}{\widetilde{\kappa}^2}$ that holds around $\omega_d=\widetilde{\omega}_r+\delta\widetilde{\omega}_r$, we can extract $P_r$ at the device. Here, $\widetilde{\kappa}$ and $\widetilde{\kappa}_{ex}$ are the total and external dissipation rate of the resonator at the dressed frequency response.
\\*

\subsubsectionnotoc{ac-voltage bias amplitude across the Josephson junctions $\widetilde{V}_{\textup{JJ}}$}

For a dispersively coupled transmon–resonator system, driving the resonator yields a voltage across the JJ of the transmon as below
\begin{equation}
\begin{split}
\label{eq_VJJ}
    \hat{V}_{\textup{JJ}} =  \frac{\hbar}{2e}\dot{\hat{\varphi}} = \frac{i}{2e}\left[ \hat{H},\hat{\varphi} \right] = \frac{4E_C}{e}(\hat{n} - n_g) - \frac{i\hbar g }{2e}(\hat{a} - \hat{a}^\dagger).
\end{split}
\end{equation}
The transmon and resonator dynamics determine the first and second term, respectively.
Under the semi-classical approximation, $i(\hat{a} - \hat{a}^\dagger) \approx ({\Omega_q}/{g})\cos\omega_d t$ is satisfied. 
Readily, we can interpret the second term in Eq.~\ref{eq_VJJ} as an ac-voltage bias across the JJ, and symbolize its amplitude by $\widetilde{V}_{\textup{JJ}}$ as in the main text.
We aim to neglect Cooper-pair breaking directly induced by the resonator field, independent of the transmon dynamics. 
Such processes would introduce substantial complexity to the system dynamics and pose significant challenges for a concise quantum-mechanical description.
Therefore, it is necessary to ensure that
\begin{equation}
\widetilde{V}_{\textup{JJ}} < \frac{2\Delta_{\textup{Al}}}{e}
\end{equation}
to avoid such complications.

\begin{figure*}
    \centering
    \includegraphics[width=0.7\linewidth]{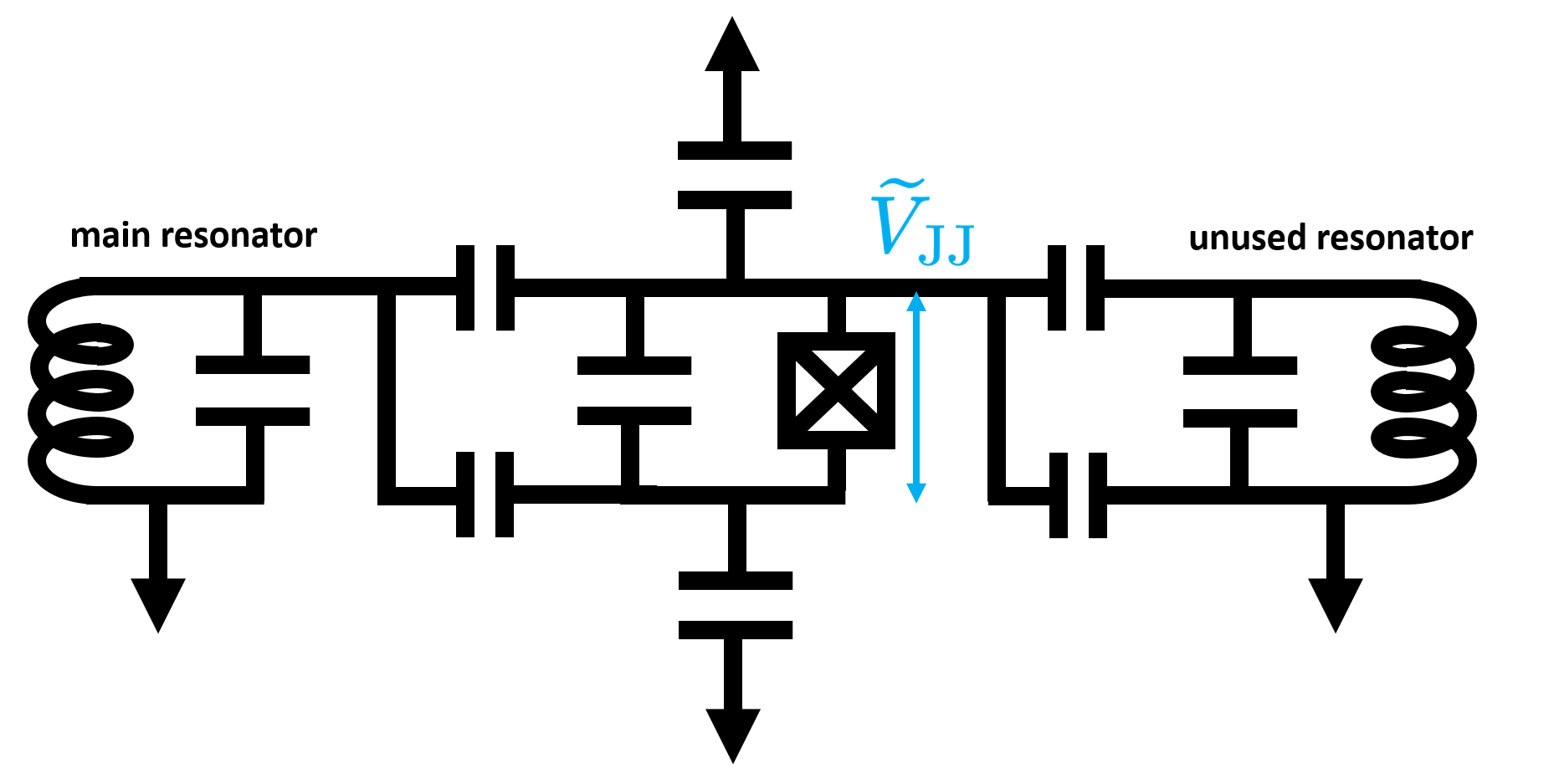}
    \caption{\textbf{Detailed circuit diagram for transmon and appended resonators.} For Q1–Q3, the real circuit images given in Fig.~\ref{fig:device} can be simplified as above. Only the fundamental modes of each co-planar resonator are presented, and they are symbolized as LC circuits. All the presented circuit elements except the JJ can be numerically estimated. Based on these information, one can estimate $\widetilde{V}_{\textup{JJ}}$, AC voltage bias accoss the JJ, for given $N_r$. For Q1-Q3, we confirm that $\widetilde{V}_{\textup{JJ}}<2\Delta_{\textup{Al}}/e$ is satisfied over the experimentally explored range. We did not estimate $\widetilde{V}_{\textup{JJ}}$ for Q4, but assume that $\widetilde{V}_{\textup{JJ}}<2\Delta_{\textup{Al}}/e$ is satisfied over the experimentally explored range.}
    \label{fig:circuit}
\end{figure*}

While $\widetilde{V}_{\textup{JJ}}$ can be roughly estimated from Eq.~\ref{eq_VJJ}, 
we incorporate additional circuit details to obtain a more accurate evaluation.
Fig.~\ref{fig:circuit} shows more detailed circuit diagram for transmon and appended resonators for Q1–Q3. Only the fundamental modes of both coplanar resonators (main and auxiliary) are taken into account, which are modeled as equivalent LC circuits. All the shown circuit parameters can be quantitatively evaluated using COMSOL Multiphysics \cite{comsol} and Quantum Circuit Analyzer Tool (QuCAT) \cite{qucat}.
Since the Josephson inductance is effectively nullified under strong resonator drives when the resonator bare response appears, we set $E_J$ to zero when estimating $\widetilde{V}_{\textup{JJ}}$.
Based on QuCAT, we find a single drive photon in the main resonator ($N_r = 1$) yields $\widetilde{V}_{\textup{JJ}}$ of 479, 495, 425 nV for Q1, Q2, and Q3, respectively. 
Using the relation $\widetilde{V}_{\textup{JJ}} \propto \sqrt{N_r}$, we can estimate $\widetilde{V}_{\textup{JJ}}$ for an arbitrary $N_r$.
For Q1–Q3, $\widetilde{V}_{\textup{JJ}}<2\Delta_{\textup{Al}}/e$ is satisfied over the experimentally explored range.
We skip estimating $\widetilde{V}_{\textup{JJ}}$ of Q4, but assume that $\widetilde{V}_{\textup{JJ}}<2\Delta_{\textup{Al}}/e$ is also satisfied over the experimentally explored range.
\\*

\subsectionntoc{Transmitted signal analysis}
\begin{figure*}
    \centering
    \includegraphics[width=0.8\linewidth]{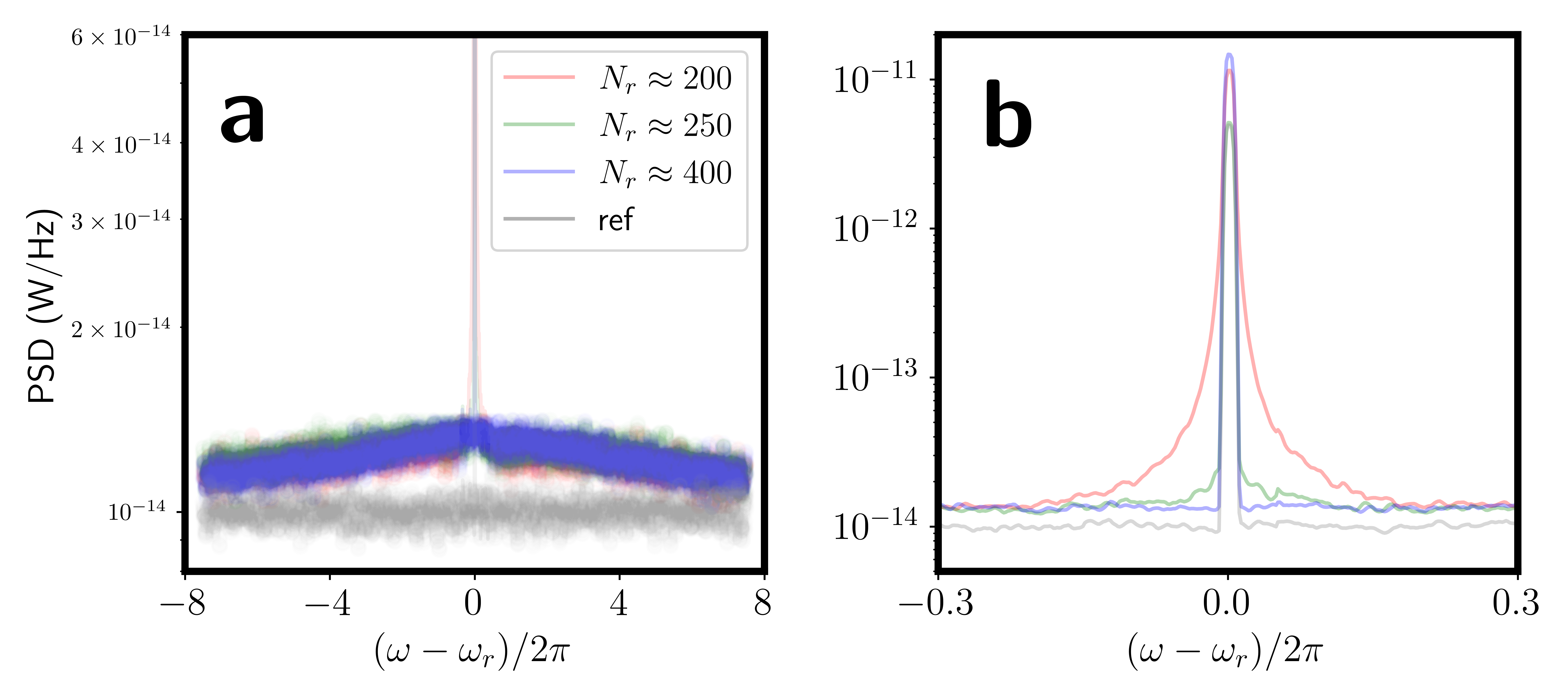}
    \caption{\textbf{Resonator output power spectra.} \textbf{(a)} Power spectral density (PSD) of the resonator output when $N_r$ is approximately 200, 250, and 400 (red, green, and blue lines). The gray line indicates the reference spectrum, when the probe field near perfectly transmits the resonator not being affected by the transmon dynamics. \textbf{(b)} Magnified view near $\omega\approx\omega_r$. Q1 is used in the presented measurement.}
    \label{fig:spectra}
\end{figure*}

Fig.~\ref{fig:spectra} displays the resonator output power spectra measured with the configuration shown in Fig.~\ref{fig:cryo}(d). 
Q1 is used in the presented measurement. 
The resonator was probed at different input powers, corresponding to estimated photon numbers $N_r \approx 200$, $250$, and $400$.
When the probe power is sufficiently low for the readout resonator to remain at its dressed frequency ($\widetilde{\omega}_r$) response regime, the signal transmits through the feedline nearly unaffected by the resonator.
The reference spectrum in Fig.~\ref{fig:spectra} is obtained under this condition, allowing us to identify how the purely coherent signal appears in the spectrum measurement.

In Fig.~\ref{fig:spectra}(a), the broad peaks and sharp spikes indicate the transmission spectra of $\delta\hat{a}$ and the coherent component of $\hat{a}$, as mentioned in the main text.
We magnify the spectra around $\omega_r$ in Fig.~\ref{fig:spectra}(b) to examine the sharp spikes. The resolution bandwidth of the spectrum analyzer is set to 10 kHz in the measurement.
A broadened feature appears for $N_r \approx 200$, which diminishes with increasing $N_r$. We attribute this feature to the resonator inhomogeneous linewidth broadening.
For $N_r \approx 400$, the spike closely matches that of the reference spectrum, implying that the linewidth broadening becomes negligible and has no significant effect on $S_{21}$.
In this regard, our assumption in the main text, neglecting the inhomogeneous linewidth broadening for $N_r \gtrsim 400$, is justified.

\clearpage

\section*{Supplementary Note 8 : Extended discussions}

\subsectionntoc{Small nonlinearity in the resonator dissipation of Q4}

In Extended Data 1, the readout resonator of Q4 exhibits nonlinear dissipation that is two orders of magnitude smaller than that of the other transmons (Q1–Q3) for a given $\Omega_q$.
Here, we clarify the origin of this reduced nonlinear dissipation.
As summarized in Table~\ref{tab:params}, Q4 has a coupling strength $g$ that is an order of magnitude smaller than those of the other transmons. Consequently, achieving the same $\Omega_q$ requires approximately two orders of magnitude larger photon number $N_r$ in the readout resonator.
For a given $\mathcal{T}$—and in the regime where $\mathcal{T}$ primarily accounts for the observed nonlinear dissipation (i.e., $\mathcal{T}/\hbar\omega_d \approx N_r(\kappa - \kappa_o)$)—we find that $\kappa - \kappa_o$ is inversely proportional to $N_r$.

Provided that the bath parameters and the structures of the chaotic Floquet modes are similar across qubits, the order of magnitude of $\mathcal{T}$ is largely determined by $\Omega_q$.
Therefore, $\kappa - \kappa_{ex}$ should scale approximately with the square of the coupling strength $g$ between the transmon and its resonator.
This reasoning explains the relatively small value of $\kappa - \kappa_{ex}$ observed for the readout resonator of Q4 at a given $\Omega_q$.
For Q4, the reduced value of $\kappa - \kappa_{ex}$ also makes the experimental data appear noisier in the plots, as the effective signal-to-noise ratio decreases.

\subsectionntoc{Experimental investigation of $\omega^c_{\textbf{QPG}}$}

Experimentally investigating $\omega^{c}_{\textbf{QPG}}$ has remained largely unexplored in studies of JJs, and—more strikingly—has not been emphasized even at the theoretical level. Our work appears to be the first to identify and characterize this cutoff dynamics in Josephson junctions.
This section illustrate the difficulty of experimental investigation of $\omega^c_{\textbf{QPG}}$. 
The challenges are non fundamental, but all practical. 

The most direct and explicit approach to determine $\omega^{c}_{\textbf{QPG}}$ would require estimating the drive voltages and QPG currents across a JJ under a time-periodic drive with frequency exceeding $\Delta/\pi\hbar$, which is approximately 80 GHz for aluminum JJs. Unfortunately, such an experiment is extremely challenging. This frequency range is unusual in conventional transport or Josephson-circuit measurements.
Moreover, accurately determining the drive voltages and currents at these high frequencies is technically demanding.
To the best of our knowledge, the experiments in Refs.~\cite{Diamond-PRXQ-2023, Liu-PRL-2024} are the only instances involving superconducting qubits where one may perform experiments with the above conditions.
Even in these works, however, the experimental signatures of non-instantaneous quasiparticle generation remain inconclusive.

In Ref.~\cite{Diamond-PRXQ-2023}, the JJ of the transmon is illuminated by incoherent radiation emitted from a resistor lamp, rather than by a coherent drive. Resolving $\omega^c_{\textbf{QPG}}$ would require accurately determining both the radiation spectrum and the power delivered to the JJ.
Accurately performing such tasks are technically challenging due to numerous practical limitations. 
Because Ref.~\cite{Diamond-PRXQ-2023} primarily investigates the flux dependence of QPG rates to distinguish between different mechanisms, precise estimation of these quantities was not essential for their analysis.
Consequently, extracting $\omega^c_{\textbf{QPG}}$ from the reported data seems not feasible.

In Ref.~\cite{Liu-PRL-2024}, the JJs of the transmons is illuminated by approximately coherent radiation with a central frequency reaching up to 500 GHz, and the authors resolve the radiation-frequency dependence of the QPG rates. The observed QPG rates qualitatively agree with their theoretical model including antenna modes of the transmons.
Although this work succeeded in resolving the antenna modes of the transmons, the main objective of the study, resolving $\omega^c_{\textbf{QPG}}$ from the experimental data seems yet unfeasible.
In the paper, the theoretical estimation on the the QPG rates are an order of magnitude smaller than the observed values.
They attribute the discrepancy to the inability to properly account for the effects of poorly characterized stray modes, which is very reasonable given the numerous technical and engineering challenges involved in accurately identifying such modes.
\\*